\documentclass[a4paper, 11pt]{article}  
\usepackage{geometry}                		
\usepackage{authblk}
\usepackage{blindtext}
\usepackage{graphicx}		
\usepackage{subcaption}		
\usepackage{amssymb}
\usepackage{amsmath}
\usepackage{algorithmic}
\usepackage{algorithm}
\usepackage{color}
\usepackage{xcolor}
\usepackage[colorlinks=true, linkcolor=blue, citecolor = blue]{hyperref}
\usepackage{booktabs}
\usepackage[toc,title]{appendix}
\newcommand{\bx}{\mathbf{x}}
\newcommand{\Cor}{\mathbb{C}\text{orr}}

\newcommand{\Var}{\mathbb{V}\text{ar}}
\newcommand{\Exp}{\mathbb{E}}
\newcommand{\ES}{\mathcal{E}_L}
\title{Cross-validation based adaptive sampling for Gaussian process models}
\author[1, 2]{Hossein Mohammadi \thanks{Corresponding Author: h.mohammadi@exeter.ac.uk}}
\author[1, 2]{Peter Challenor}
\author[1]{Daniel Williamson}
\author[1, 2]{Marc Goodfellow}
\affil[1]{College of Engineering, Mathematics and Physical Sciences, University of Exeter, Exeter, UK}
\affil[2]{EPSRC Centre for Predictive Modelling in Healthcare, University of Exeter, Exeter, UK}
\date{}							
\begin{document}
\maketitle
\begin{abstract}
In many real-world applications, we are interested in approximating black-box, costly functions as accurately as possible with the smallest number of function evaluations. A complex computer code is an example of such a function. In this work, a Gaussian process (GP) emulator is used to approximate the output of complex computer code. We consider the problem of extending an initial experiment (set of model runs) sequentially to improve the emulator.
A sequential sampling approach based on leave-one-out (LOO) cross-validation is proposed that can be easily extended to a batch mode. This is a desirable property since it saves the user time when parallel computing is available. After fitting a GP to training data points, the expected squared LOO (ES-LOO) error is calculated at each design point. ES-LOO is used as a measure to identify important data points. More precisely, when this quantity is large at a point it means that the quality of prediction depends a great deal on that point and adding more samples nearby could improve the accuracy of the GP. As a result, it is reasonable to select the next sample where ES-LOO is maximised. However, ES-LOO is only known at the experimental design and needs to be estimated at unobserved points. To do this, a second GP is fitted to the ES-LOO errors and  where the maximum of the modified expected improvement (EI) criterion occurs is chosen as the next sample. EI is a popular acquisition function in Bayesian optimisation and is used to trade-off between local/global search. However, it has a tendency towards exploitation, meaning that its maximum is close to the (current) ``best" sample. To avoid clustering, a modified version of EI, called pseudo expected improvement, is employed which is more explorative than EI yet allows us to discover unexplored regions. Our results show that the proposed sampling method is promising. 
\end{abstract}
\section{Introduction}
\label{sec:introduction}
In many real-world applications, we are interested in predicting the output of complex computer models (or simulators) such as high fidelity numerical solvers. The reason is that such models are computationally intensive and we cannot use them to perform analysis that requires very many runs. One way to predict the model output is to use surrogate models also known as emulators which are constructed based on a limited number of simulation runs. Surrogates are fast to run and the analysis can be carried out on them, see e.g. \cite{liu2017, vernon2018, banyay2019}. Among different classes of surrogate models, Gaussian process (GP) emulators \cite{GPML} have gained increasing attention due to their statistical properties such as computational tractability and flexibility. GPs provide a flexible paradigm to approximate any smooth, continuous function \cite{neal1998} thanks to the variety of covariance kernels available. Most importantly, the GP prediction is equipped with an estimation of uncertainty which reflects the accuracy of the prediction. 

One factor that heavily affects the accuracy of emulators is the location of the training data, also called the design, \cite{simpson2001, crombecq2011}. In this context, the design of computer experiments has become an integral part of the analysis of computer experiments \cite{sacks1989, santner2003}. Generally speaking, such design can be performed in a \emph{one-shot} or \emph{adaptive} manner \cite{liu2018}. In the former all samples are chosen at once while in the latter the points are selected sequentially using information from the emulator and the existing data. Examples of one-shot design of experiments (DoEs) methods are the Latin hypercube \cite{mckay1979}, full factorial \cite{box1961}, orthogonal array \cite{owen1992}, minimax and maximin-distance designs \cite{johnson1990}. 
A potential drawback of one-shot DoEs is that they may result in under/oversampling and can waste computational resources \cite{sheikholeslami2017, garud2017}. However, this is not the case for adaptive approaches where we can stop the computationally expensive sampling process as soon as the emulator reaches an acceptable level of accuracy. Moreover, with adaptive sampling it is possible to take more samples in ``interesting" regions where, e.g., the underlying function is highly nonlinear or exhibits abrupt changes. This paper focuses on GP-based adaptive sampling where an initial design is extended sequentially to improve the emulator. The initial DoE is often \emph{space-filling} meaning that the points are scattered uniformly over the input space. We refer the reader to \cite{pronzato2012, joseph2016} and references therein for more information on space-filling designs.

There are various GP-based adaptive sampling methods which can be categorised according to their selection criteria, i.e. the strategies to find future designs. The readers are referred to \cite{garud2017, liu2018} for a comprehensive review of the existing methods. An intuitive criterion is the built-in predictive variance of GPs, also known as the prediction uncertainty or mean squared error (MSE). The idea is that the predictive variance is regarded as an estimation of the ``real" prediction error and a point with the maximum uncertainty is taken as the next experimental design \cite{martin2002, jin2002}. The predictive variance increases away from the data points. It is highly probable that sampling based on the MSE criterion gives some sort of space-filling design which can be achieved using one-shot techniques. Note that the MSE criterion (see \ref{E:kriging_cov}) depends only on the location of samples and not the output values. Thus, the MSE-based sampling strategy can be regarded as non-adaptive in the sense that it does not consider output information. Moreover, MSE is large on the boundaries of the input space and that can lead to taking a lot of samples on the boundaries. However, this is not desirable in many situations especially when the main characteristics of the true function appear inside the interior region and when the dimension of the input space is high. The integrated mean square error (IMSE) is a variant of the MSE criterion and selects a new point if adding that point to the existing design minimises the integral of the MSE \cite{sacks1989, picheny2010}. However, computing IMSE can be cumbersome, especially in high dimensions.

Maximum entropy is another common selection criterion in the adaptive sampling paradigm \cite{shewry1987, koehler1996}. It is equivalent to the maximum MSE criteria under certain circumstances \cite{jin2002, lam2008}. As a result, an adaptive sampling strategy based on maximum entropy tends to place many points at the borders of the input space \cite{koehler1996}. This issue can be mitigated using \emph{mutual information} (MI) as proposed by Krause et al. \cite{krause2008} in the sensor placement problem. The MI of two random variables is a measure of reduction in the uncertainty of one random variable through observing the other one. A sequential design approach is developed in \cite{beck2016} where the MI criterion is modified by introducing an extra parameter, called \emph{nugget}, to the correlation matrix of the GP (in the denominator), see \ref{MICE}. The inclusion of the nugget parameter prevents selecting a new sample close to the current design. The algorithm called MICE (mutual information for computer experiments) is then used to emulate a simulator.

The leave-one-out (LOO) cross-validation (CV) error defines another class of adaptive sampling criterion \cite{li2009, legratiet2012, aute2013, liu2015}. To obtain the LOO error at an experimental design we remove that point from the training data set. Then, a GP is fitted to the remaining samples and the response at the left out point is predicted. The difference between the predicted and actual response serves as the LOO error. A relatively small error indicates that the prediction accuracy in a vicinity of the removed point is high and there is good information about the true function there. On the other hand, a comparably large error means that the removed point has a huge impact on the accuracy of the emulator and hence, we need more samples in the nearby region to reduce the errors. In \cite{volodina2020} scores obtained by CV are used to identify regions of distinct model behaviour and specify mixture of covariance functions for GP emulators. Using the LOO errors as a sampling criterion has several advantages. First, it provides actual prediction error at the design points. Second, it is model-independent and can be achieved by any surrogate model. For example, in \cite{bensalem2017} a methodology based on the LOO CV is proposed to estimate the prediction uncertainty of any surrogate model, either deterministic or probabilistic. Third, computing the LOO errors is not expensive in terms of computational cost \cite{dubrule1983}. However, the LOO errors are not determined everywhere in the input space and only defined at locations where we have pre-existing model runs.

It is worth mentioning that CV is used for other purposes (such as model selection/fitting, diagnostic and parameter inference) than adaptive sampling. A survey of CV strategies on the model selection can be found in \cite{alort2010, zhang2015}. \cite{ginsbourger2021} suggested an efficient (multiple-fold) CV expression for GP model fitting and diagnostics. The CV technique is employed by \cite{maatouk2015} to estimate the covariance parameters of a GP with inequality constraints. Bachoc \cite{bachoc2013} studied the capability of CV (and maximum likelihood) in estimating the parameters of a GP with a misspecified covariance structure. A probabilistic version of CV is proposed in \cite{martino2017} and is reported robust against mismatch between the data and chosen model. Viana et al. \cite{viana2009} proposed to predict computer codes by an ensemble of surrogate models such that CV serves as a performance measure of surrogates to select the best one.  \cite{liang2014} suggested to use CV for the optimal basis functions selection in designing the GP (prior) mean. A methodology for the Bayesian time series analysis is presented in \cite{bukner2020} where the forecast of future observations is performed via CV. In \cite{youngmok2014} the human gait pattern kinematics is predicted with a GP regression whose validation is done by CV. 

This paper proposes an adaptive sampling DoE relying on the LOO cross-validation method to build GP emulators as accurately as possible for deterministic computer codes over the entire domain. The proposed method has a few parameters to be tuned and can be extended to a batch mode where at each iteration a set of inputs is selected for evaluation. This is an important property as it saves the user time when parallel computing is available \cite{williamson2015}. The remainder of the paper is organised as follows. In the next section, the statistical methodology of GP emulators is briefly reviewed. \ref{sec:adaptive_sampling} introduces the proposed adaptive sampling approach and its extension to batch mode. \ref{sec:experiments} presents numerical experiments where the predictive performance of our algorithm is tested. Finally, the paper's conclusion is in \ref{sec:conclusion}.
\section{Gaussian process models}
\label{sec:GP}
First we look at GP emulators and their statistical background. Let the underlying function of a deterministic complex computer code be given by $f: \mathcal D \mapsto \mathbb{R}$ in which $\mathcal D$ is a compact set in $\mathbb R^d$. Suppose $\mathcal{A} = \{\mathbf{X}_n,  \mathbf{y}_n\}$ is a training data set where $\mathbf{X}_n=\left(\bx_1, \ldots, \bx_n \right)^\top$ and $\mathbf{y}_n =\left(f(\bx_1), \dots, f(\bx_n) \right)^\top$ represent $n$ locations in the input space $\mathcal D$ and the corresponding outputs (responses/observations), respectively. Let $\left(Z_0(\bx)\right)_{\bx \in \mathcal{D}}$ be the Gaussian process by which we want to model $f$. In this framework, it is assumed that $\mathbf{y}_n$ has a multivariate normal distribution given by 
\begin{equation}
	\mathbf{y}_n \sim \mathcal{N} \left(m_0(\mathbf{X}_n),  k_0(\mathbf{X}_n, \mathbf{X}_n)\right) ,
\end{equation}
where $m_0$ and $k_0$ are the (preselected) mean and covariance functions of $\left(Z_0(\bx)\right)_{\bx \in \mathcal{D}}$. Without loss of generality, we assume that the mean function is a constant: $m_0(x) = \mu$.  The positive semi-definite covariance function $k_0$ plays an important role in GP modelling; assumptions about the underlying function such as differentiability or periodicity are encoded through kernels. The Mat\'ern family of kernels are commonplace in computer experiments and (in the univariate case) are defined as
\begin{equation}
	k_0(x, x^\prime) = \sigma^2 \frac{2^{1 - \nu}}{\Gamma(\nu)} \left(\frac{\sqrt{2\nu}}{\theta} |x - x^\prime| \right)^\nu B_\nu \left( \frac{\sqrt{2\nu}}{\theta} |x - x^\prime| \right) , 
	\label{Mat_kernel}
\end{equation}
where  $\Gamma(\cdot)$ is the Gamma function and $B_\nu(\cdot)$ denotes the modified Bessel function of the second kind of order $\nu$. The parameter $\nu$ regulates the degree of smoothness of the GP sample paths/functions such that a process with the Mat\'ern kernel of order $\nu$ is $\lceil \nu -1 \rceil$ times differentiable \cite{GPML}. The positive parameters $\sigma^2$ and $\theta$ are referred to as the \emph{process variance} and \emph{correlation length scale}, respectively. These parameters are usually unknown and need to be estimated from data. We refer the reader to \cite{GPML, BayesianDataAnalysis} for a more detailed explanation on parameter estimation techniques. 

Given that the unknown parameters are estimated, the posterior predictive distribution relying on $Z_n(\bx) = Z_0(\bx) \mid \mathcal{A}$ can be calculated. The posterior mean and covariance at a generic location $\bx$ have closed-form expressions as \cite{GPML}
\begin{flalign} 
	m_n(\bx) & =  \hat{\mu} + \mathbf{k}(\bx)^\top \mathbf{K}^{-1}(\mathbf{y}_n -  \hat{\mu} \mathbf{1})  \label{E:kriging_mean}, \\
	\nonumber k_n(\bx, \bx^\prime) & = k_0(\bx, \bx^\prime)  - \mathbf{k}(\mathbf{x})^\top \mathbf{K}^{-1}\textbf{k}(\mathbf{x^\prime}) \\ 
	& + \frac{\left(1 - \mathbf{1}^\top\mathbf{K}^{-1}\mathbf{k}(\bx)\right)\left(1 - \mathbf{1}^\top\mathbf{K}^{-1}\mathbf{k}(\bx^\prime)\right)}{\mathbf{1}^\top\mathbf{K}^{-1}\mathbf{1}} .
	\label{E:kriging_cov}
\end{flalign}
Here, $\textbf{k}(\textbf{x})= \left(k_0(\bx, \bx_1), \dots, k_0(\bx, \bx_n)\right)^\top$ is the vector of covariances between $Z_0(\bx)$ and $Z_0(\bx_i)$s, $\mathbf{1}$ is a vector of ones and $\textbf{K}$ is an $n \times n$ covariance matrix with elements $\textbf{K}_{ij} = k_0(\bx_i, \bx_j),~ \forall \, 1\leq i, j \leq n$. The last term in \ref{E:kriging_cov} reflects the additional uncertainty concerning the estimation of $\mu$. We refer the reader to  \cite{roustant2012} where the ``ordinary kriging" is described. 
The predictive variance $s_n^2(\bx) =  k_n \left(\bx, \bx \right)$ determines the uncertainty associated with the prediction at $\bx\in \mathcal{D}$ and is a measure of the prediction accuracy. In adaptive sampling based on the MSE criterion, the new sample is the point that maximises $s_n^2(\bx)$.
\section{Proposed adaptive sampling method}
\label{sec:adaptive_sampling}
 Before introducing our method, it is worth mentioning that an ``efficient" adaptive sampling strategy should meet the conditions below \cite{liu2018}.
 \begin{itemize}
 	\item[(i)] \textbf{Local exploitation:} that allows us to add more points in interesting areas discovered so far.    
 	\item[(ii)]  \textbf{Global exploration:} by which unexplored domain regions can be detected. 
 	\item[(iii)]  \textbf{Trade off between local exploitation and global exploration:} which balances the previous two objectives using a suitable measure.
 \end{itemize}   
 The proposed adaptive sampling approach uses \emph{expected squared LOO error}, see \ref{sec:squar_LOO_error}, for local exploitation. The global exploration and the trade off between the local/global search is driven by the \emph{pseudo expected improvement} criterion introduced in \ref{sec:PEI}. 
\subsection{Expected squared LOO cross-validation error}
\label{sec:squar_LOO_error}
We wish to improve the GP emulator $Z_n(\bx)$ constructed on the training data set $\mathcal{A}$ using the LOO CV method. To do so, the first step is to obtain the LOO errors. Let $e_{L}(\bx_i)$ denote the LOO CV error at the design sites $\bx_i, i = 1, \ldots,  n$. The computation of $e_{L}(\bx_i)$ relies on the Gaussian process $Z_{n, -i}(\bx)$ which is obtained by conditioning $Z_0(\bx)$ on all observations except the $i$-th one:
$Z_{n, -i}(\bx) = Z_0(\bx) \mid \,  \mathbf{y}_n \setminus \{f(\bx_i)\} $. The predictive mean and variance of $Z_{n, -i}(\bx)$ are shown by $m_{n, -i}(\bx)$ and $s^2_{n, -i}(\bx)$, respectively. The LOO CV error is then calculated as
\begin{equation} 
	e_{L}(\bx_i) = \lvert m_{n, -i}(\bx_i) - f(\bx_i) \rvert ,
	\label{LOO_error}
\end{equation}
which can be regarded as the  sensitivity of the emulator to the left out point $\bx_i$. As a result, the idea of adding new samples near the points with large LOO errors is used in several adaptive sampling works, see e.g. \cite{li2009, aute2013}. In this work, the unknown parameters of $Z_n(\bx)$ are estimated at each iteration when a new point is added to the existing data. However, we do not estimate the parameters of $Z_{n, -i}(\bx)$ to alleviate the computational burden. To this end, the estimated parameters of $Z_n(\bx)$ are used in $Z_{n, -i}(\bx)$. Moreover, \ref{LOO_error} can be calculated efficiently using the formula proposed in \cite{dubrule1983}.

\begin{figure}[htpb] 
	\includegraphics[width=0.48\textwidth]{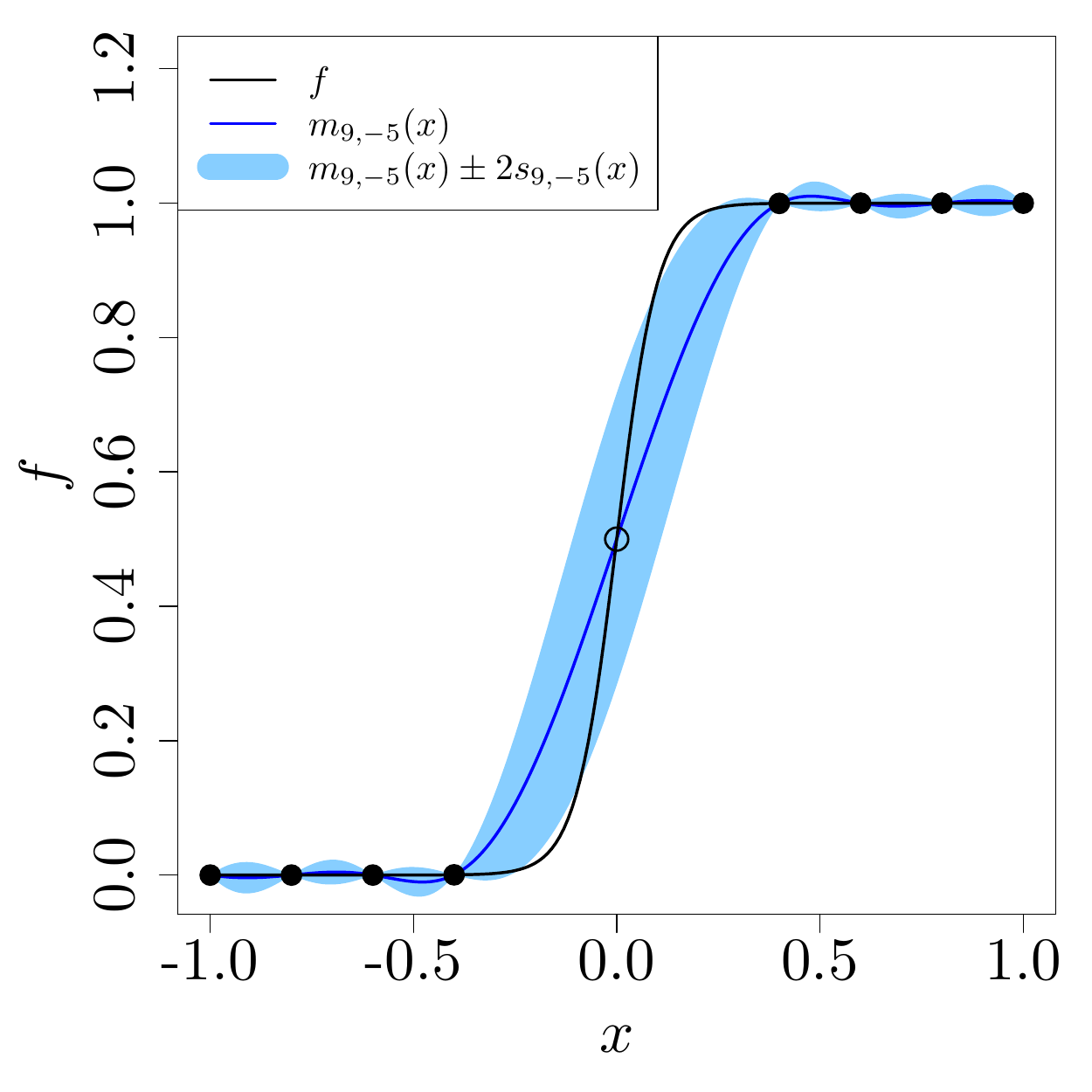}
	\includegraphics[width=0.48\textwidth]{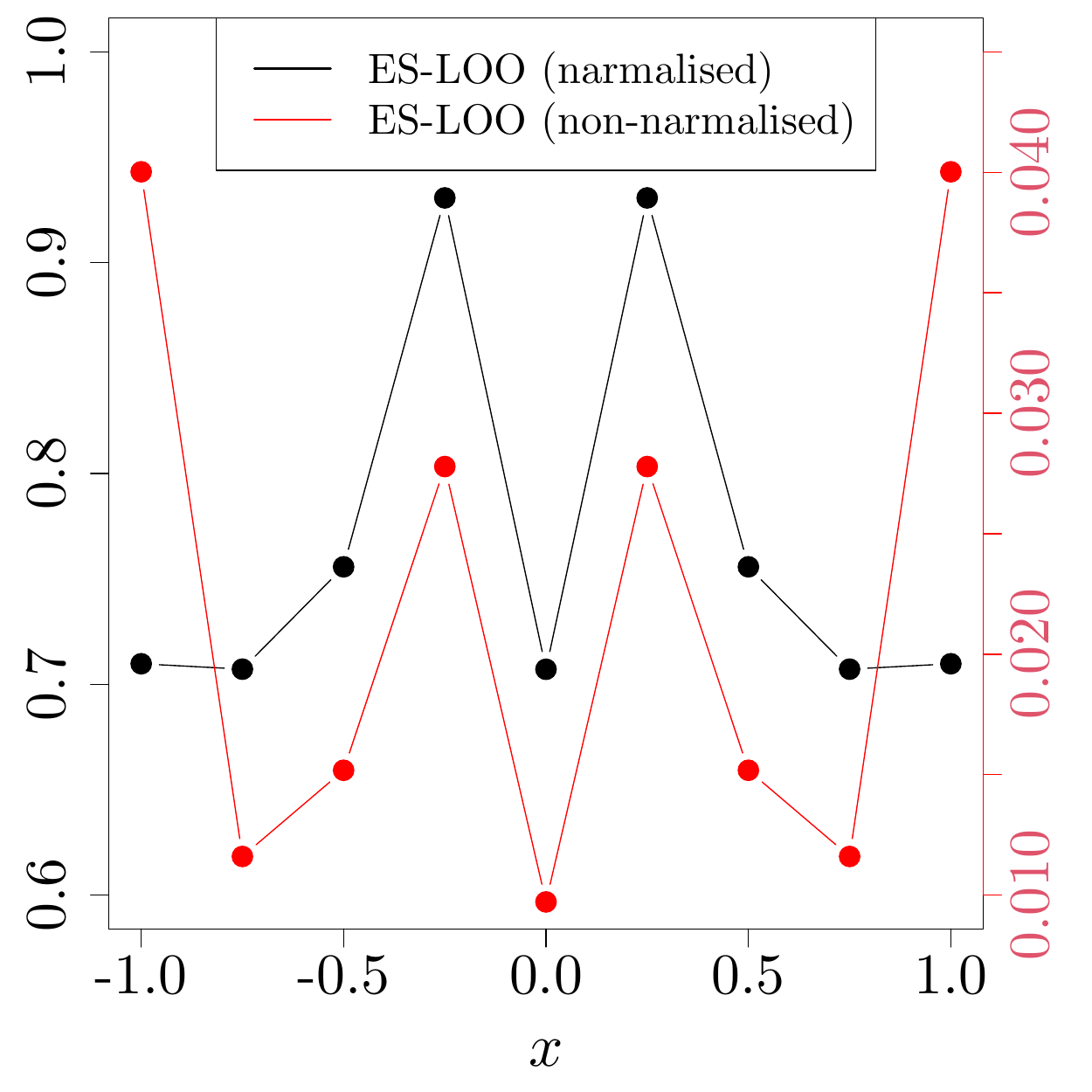}
	\caption{Left: The LOO error (\ref{LOO_error}) can be misleading as a selection criterion. Removing the fifth sample, i.e. $(0, 0.5)$, does not change the prediction at $x = 0$ and the LOO error remains zero there: $\lvert m_{9, -5}(x_5) - f(x_5) \rvert = 0$.  This means that the point $(0, 0.5)$ does not have any influence on the emulator while $f$ has a large gradient there. However, the ES-LOO error which accounts for the prediction uncertainty is not zero at $x_5$.  Right: Normalised (black) vs non-normalised (red) ES-LOO errors for the sigmoid function visualised on the left panel. The value of non-normalised ES-LOO is maximum at the endpoints where adding new samples does not improve the emulator. The true function is $f(x) = \frac{1}{1 + \exp(-20x)}$.}
	\label{ExpLOO_vs_LOO}
\end{figure}

The LOO error $e_{L}(\bx_i)$ only accounts for the difference between the predictive mean and the real value at $\bx_i$ and can be a misleading selection criterion in some situations. \ref{ExpLOO_vs_LOO} (left panel) shows an example where the prediction at $x_5 = 0$ is equal to the true value there and therefore $e_{L}(x_5) = 0$. This means that the chance of adding a new sample near the fifth data point is low although it is in a crucial region. The above mentioned problem can be mitigated using ``expected squared LOO error (ES-LOO)" which provides more information than the LOO error about the sensitivity of the emulator to the design points. More precisely, ES-LOO accounts for both the prediction uncertainty and the difference between the prediction and the true value, as described below.

Let $\ES(\bx_i)$ represent the value of (normalised) ES-LOO error at the design site $\bx_i$. It is defined as
\begin{equation}
	\ES(\bx_i)  = \frac{\Exp \left[ \left( Z_{n, -i}(\bx_i) - f(\bx_i) \right)^2 \right]}{\sqrt{\Var\left( \left( Z_{n, -i}(\bx_i) - f(\bx_i) \right)^2 \right) }} ,
	\label{exp_squar_LOO}
\end{equation}
where
\begin{align}
	\label{Expected_squar_LOO}
	\Exp \left[ \left( Z_{n, -i}(\bx_i) - f(\bx_i) \right)^2 \right] &= s^2_{n, -i}(\bx_i) + \left( m_{n, -i}(\bx_i) - f(\bx_i)  \right)^2 ,  \\
	\label{Var_squar_LOO}
	\Var\left( \left( Z_{n, -i}(\bx_i) - f(\bx_i) \right)^2 \right) &= 2s^4_{n, -i}(\bx_i) + 4 s^2_{n, -i}(\bx_i)  \left( m_{n, -i}(\bx_i) - f(\bx_i)  \right)^2 .
\end{align} 
The latter is used in \ref{exp_squar_LOO} to normalise ES-LOO values. It is recommended in \cite{legratiet2012, legratiet2015} that standardizing the LOO errors yields a better measure for adaptive sampling. Our experiments also suggest that normalised ES-LOO should be preferred over non-normalised one. For example, in \ref{ExpLOO_vs_LOO} (right panel) the normalised (black) and non-normalised (red) ES-LOO are shown for the sigmoid function. As can be seen, the non-normalised ES-LOO error is maximum at the endpoints where adding new samples does not improve the accuracy of the emulator significantly. In the standardized case, however, the value of ES-LOO is moderated at the endpoints since the term $s_{n, -i}(\bx_i)$ plays an important role in \ref{Var_squar_LOO}.

\begin{figure}[htpb] 
	\includegraphics[width=0.49\textwidth]{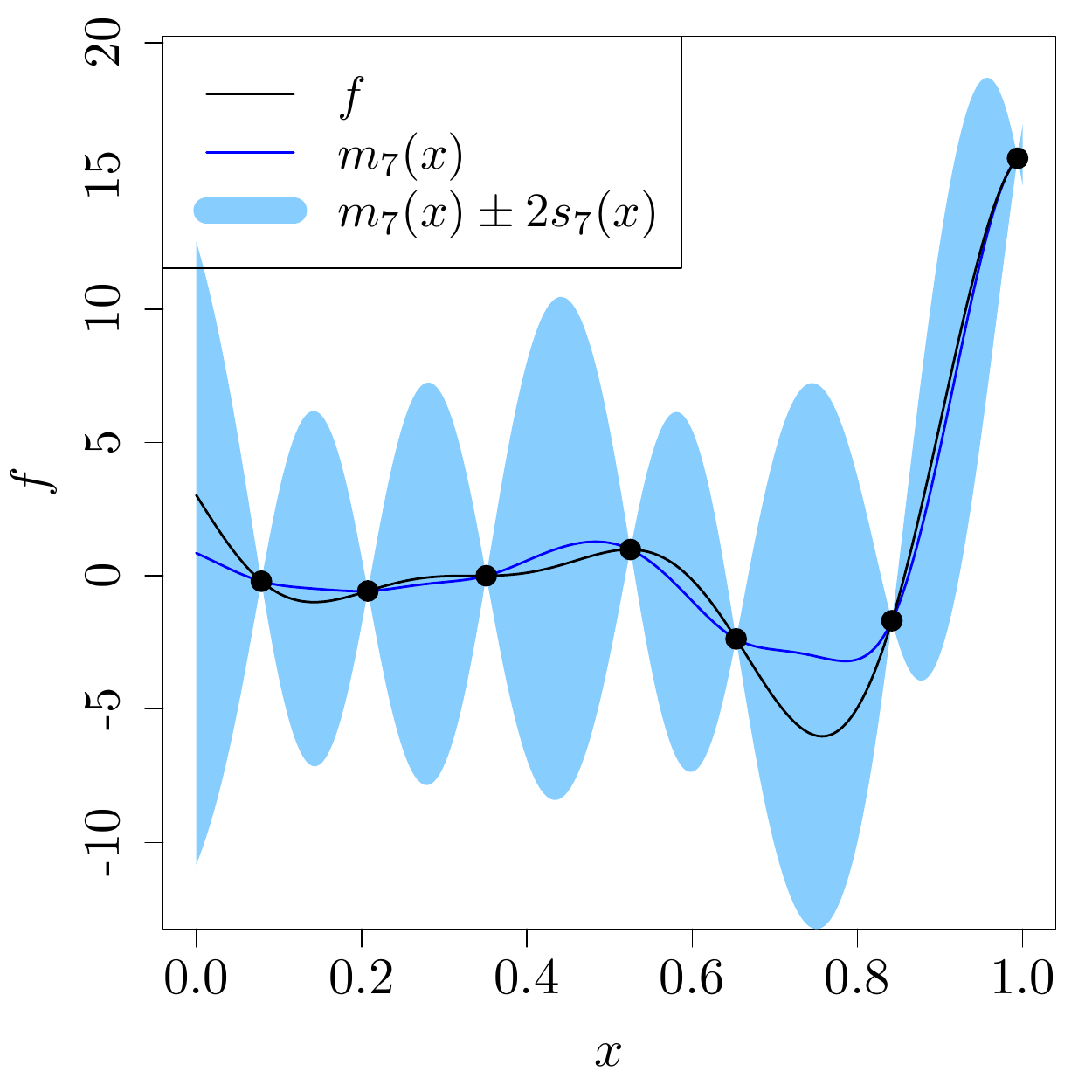}
	\includegraphics[width=0.49\textwidth]{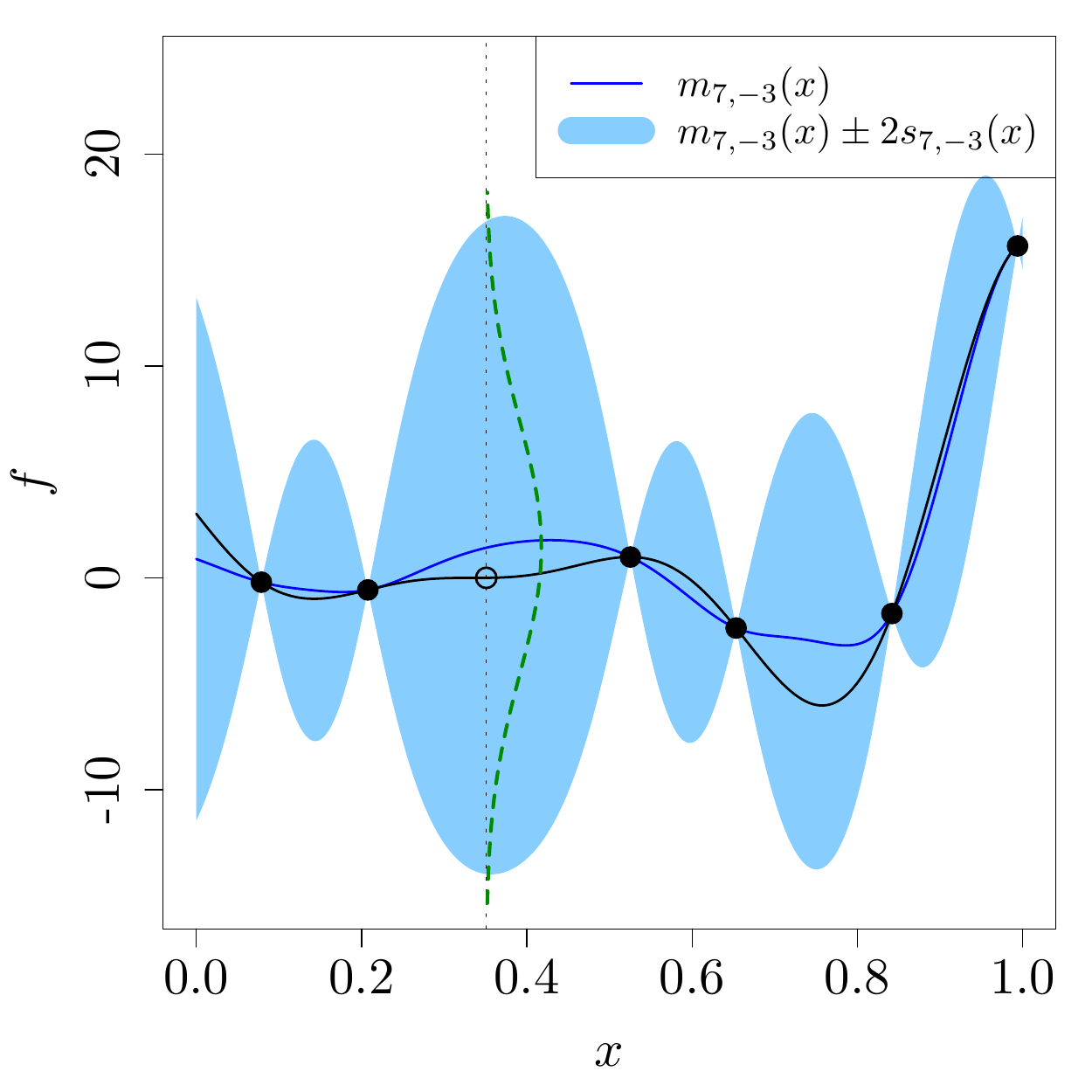}
	\caption{Left: Gaussian process prediction (blue) with $95\%$ credible intervals (shaded) based on 7 observations from the true function (black). Right: GP prediction based on the training data except the third one where the GP has a normal distribution specified by \ref{normal_dist_i}.}
	\label{GP_CV}
\end{figure}

To see how \ref{Expected_squar_LOO} and \ref{Var_squar_LOO} are obtained we first note  that
\begin{equation}
	Z_{n, -i}(\bx_i) \sim \mathcal{N} \left(m_{n, -i}(\bx_i),  s^2_{n, -i}(\bx_i) \right) ,
	\label{normal_dist_i}
\end{equation}
as is shown with an example in \ref{GP_CV}. By standardizing the above equation we reach
\begin{equation}
	\frac{Z_{n, -i}(\bx_i) - f(\bx_i)}{s_{n, -i}(\bx_i)} \sim \mathcal{N} \left(\frac{m_{n, -i}(\bx_i)  - f(\bx_i)}{s_{n, -i}(\bx_i)} ,  1 \right) ,
	\label{standard_normal_dist_i}
\end{equation}
in which the square of the left hand side is a random variable with noncentral chi-square distribution characterised by 
\begin{equation}
	\left( \frac{Z_{n, -i}(\bx_i) - f(\bx_i)}{s_{n, -i}(\bx_i)} \right)^2 \sim \chi^{\prime}{^2} \left(\kappa = 1, \lambda = \left( \frac{m_{n, -i}(\bx_i)  - f(\bx_i)}{s_{n, -i}(\bx_i)} \right)^2 \right) .
\end{equation}
Here, $\kappa$ and $\lambda$ are the degrees of freedom and noncentrality parameter, respectively\footnote{Suppose $X_1, \ldots, X_\kappa$ are $\kappa$ independent random normal variables such that $X_i \sim \mathcal{N} \left(\mu_i, 1 \right) , 1 \leq i \leq \kappa$. Then, $\sum_{i = 1}^{\kappa} X_i^2 \sim \chi^{\prime}{^2} \left(\kappa, \lambda = \sum_{i = 1}^{\kappa} \mu_i^2 \right)$ has a noncentral chi-square distribution with mean $\kappa + \lambda$ and variance $2\left(\kappa + 2\lambda \right)$.}. As a result
\begin{align}
	\Exp \left[ \left( \frac{Z_{n, -i}(\bx_i) - f(\bx_i)}{s_{n, -i}(\bx_i)} \right)^2 \right] &= 1 + \left( \frac{m_{n, -i}(\bx_i) - f(\bx_i)}{s_{n, -i}(\bx_i)} \right)^2 , \\
	\Var \left( \left( \frac{Z_{n, -i}(\bx_i) - f(\bx_i)}{s_{n, -i}(\bx_i)} \right)^2 \right) &= 2\left(1 + 2\left( \frac{m_{n, -i}(\bx_i) - f(\bx_i)}{s_{n, -i}(\bx_i)} \right)^2  \right) .
\end{align}  
Finally, if the expectation and variance in the above equations are multiplied by $s^2_{n, -i}(\bx_i)$ and $s^4_{n, -i}(\bx_i)$ respectively, we reach \ref{Expected_squar_LOO} and \ref{Var_squar_LOO}.

The analytical expression of \ref{Expected_squar_LOO} is similar to the \emph{expected improvement for global fit (EIGF)} infill sampling criterion proposed by Lam \cite{lam2008}. In \ref{sec:experiments}, we compare the predictive performance of our method with EIGF on several test functions. In EIGF, the improvement at an arbitrary point $\bx$ is given by:
\begin{equation}
	IGF(\bx) = \left( Z_n(\bx) - f(\bx_i^*) \right)^2,
	\label{IGF}
\end{equation}   
where $f(\bx_i^*)$ is the response at the location $\bx_i^*$ which is closest
(in Euclidean distance) to $\bx$. The EIGF criterion is the expected value of $IGF(\bx)$ and takes the following form
\begin{equation}
	EIGF(\bx) = \Exp [IGF(\bx)] = \left( m_n(\bx) - f(\bx_i^*) \right)^2 + s_n^2(\bx) .
	\label{EIGF_criterion}
\end{equation}
Using EIGF as the selection criterion, the next sample is chosen where EIGF is maximum: 
\begin{equation*}
	\bx_{n+1} = \underset{\bx \in \mathcal{D}}{\arg\!\max}~ EIGF(\bx).
\end{equation*}
\subsection{Proposed selection criterion}
\label{sec:PEI}
Since the magnitude of $\ES(\bx_i)$ reflects the sensitivity of the emulator to the loss of information provided by the function evaluation at $\bx_i$, it is reasonable to choose the next sample where ES-LOO is maximum. However, this quantity is only defined at the training data while we need to look for the next design point out-of-sample. In this work, we extend the ES-LOO error to be a function defined over the whole domain, $\ES(\bx), \bx \in \mathcal{D}$, that we have observed at the design points, and we model this function with a GP. The interpretation of  $\ES(\bx)$ is the value of the ES-LOO error we think we would see if $(\bx, f(\bx))$ were part of our data set. The GP model to estimate $\ES(\bx)$ at unobserved locations is denoted by $Z_n^e(\bx)$ whose predictive mean and variance are indicated by $m_n^{e}(\bx)$ and $s_n^{e}(\bx)$, respectively. The training set for the second GP is $\lbrace \mathbf{X}_n, \mathbf{y}_n^e  \rbrace$ where $\mathbf{y}_n^e = \left(\ES(\bx_1), \ldots, \ES(\bx_n) \right)^\top$.

\begin{figure}[htpb] 
	\centering
	\includegraphics[width=0.5\textwidth]{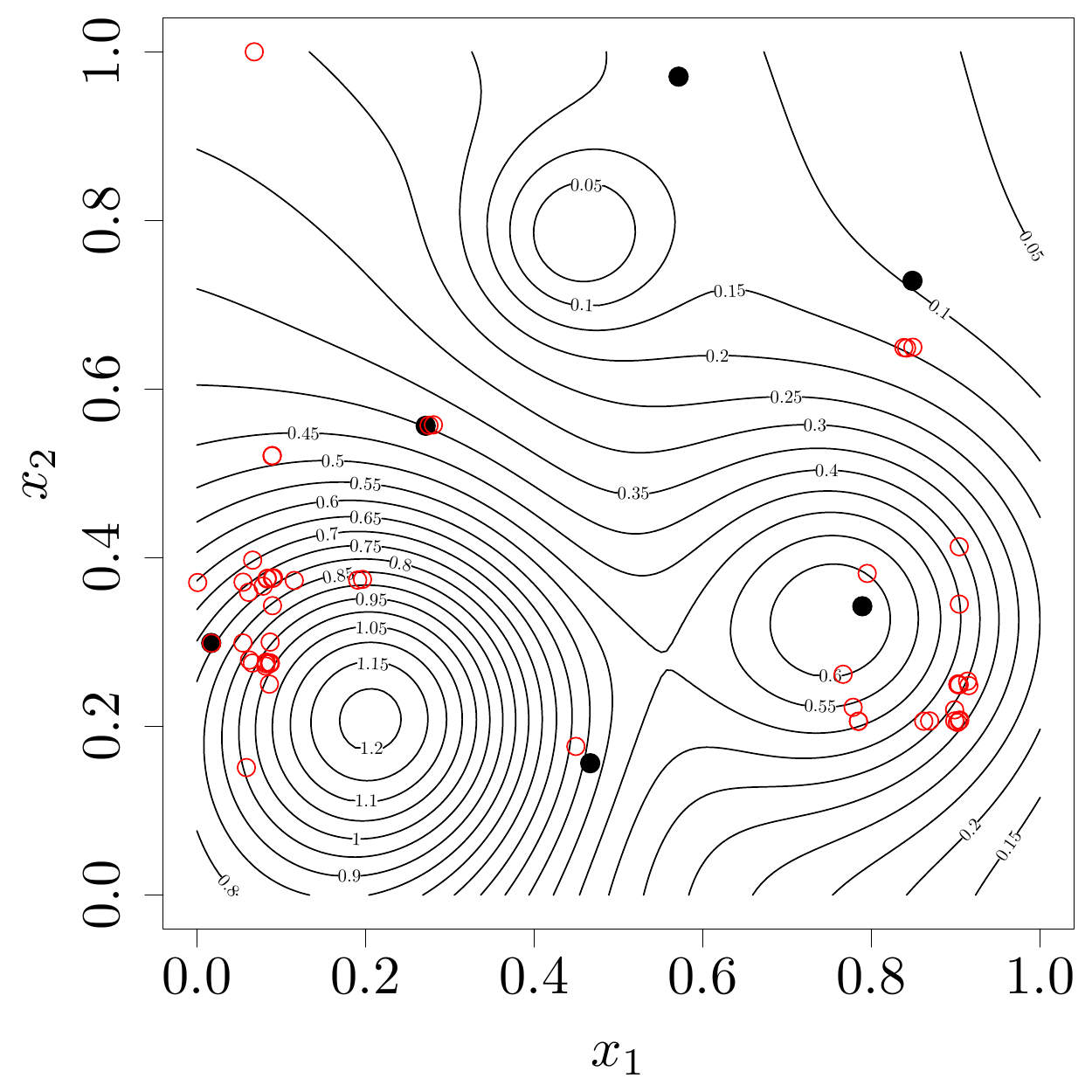}
	\caption{Adaptive designs (red circles) are obtained by maximising $\ES(\bx)$ using the EI criterion. The filled circles are the initial design and the true function is Franke's function. EI tends towards exploitation and as a result clustering occurs.}
	\label{franke_EI}
\end{figure}

After estimating $\ES(\bx)$, we can find its maximum applying techniques in surrogate-based optimisation. In this framework, a naive approach is to maximise $m_n^{e}(\bx)$. However, this simple strategy does not define a valid optimisation scheme due to \emph{overexploitation} \cite{jones2001} meaning that the new samples are taken very close to the points with a large ES-LOO error. To overcome this problem, we need to take into account $s_n^{e}(\bx)$ as the exploration component in the course of the search. To this end, we employ \emph{Expected improvement (EI)} which is one of the most common acquisition functions in Bayesian optimisation \cite{brochu2010, jones2001} to trade-off between exploration and exploitation. It is expressed via
\begin{align} 
	EI(\bx) &=\begin{cases}
		\left(m_n^{e}(\bx) -\max (\mathbf{y}_n^e)\right) \Phi(u) + s_n^{e}(\bx) \phi(u)  & \text{if ~ $s_n^{e}(\bx) > 0$}\\
		0 & \text{if ~ $s_n^{e}(\bx) = 0$} ,
	\end{cases}
	\label{ei_criterion}
\end{align}
where $u = \frac{m_n^{e}(\bx) - \max(\mathbf{y}_n^e)}{s_n^{e}(\bx)}$ and $\phi(\cdot)$ and $\Phi(\cdot)$ represent the PDF and CDF of the standard normal distribution, respectively. EI is a non-negative, parameter-free function and is zero at the data points. However, it is shown that EI is biased towards exploitation especially at the beginning of the search \cite{schonlau1997, jones1998, ponweiser2008}. As a result, if EI is used to find the maximum of $\ES(\bx)$ at each iteration, the new samples are clustered. This is illustrated by an example in \ref{franke_EI} where the black circles are the initial design and the red circles represent the new samples selected based on the EI criterion. Clusters of the new points can be detected on the contour plot of Franke's function (its analytic expression is given in \ref{toy_tests}) due to the tendency of EI towards exploitation.

\begin{figure}[htpb] 
	\includegraphics[width=0.49\textwidth]{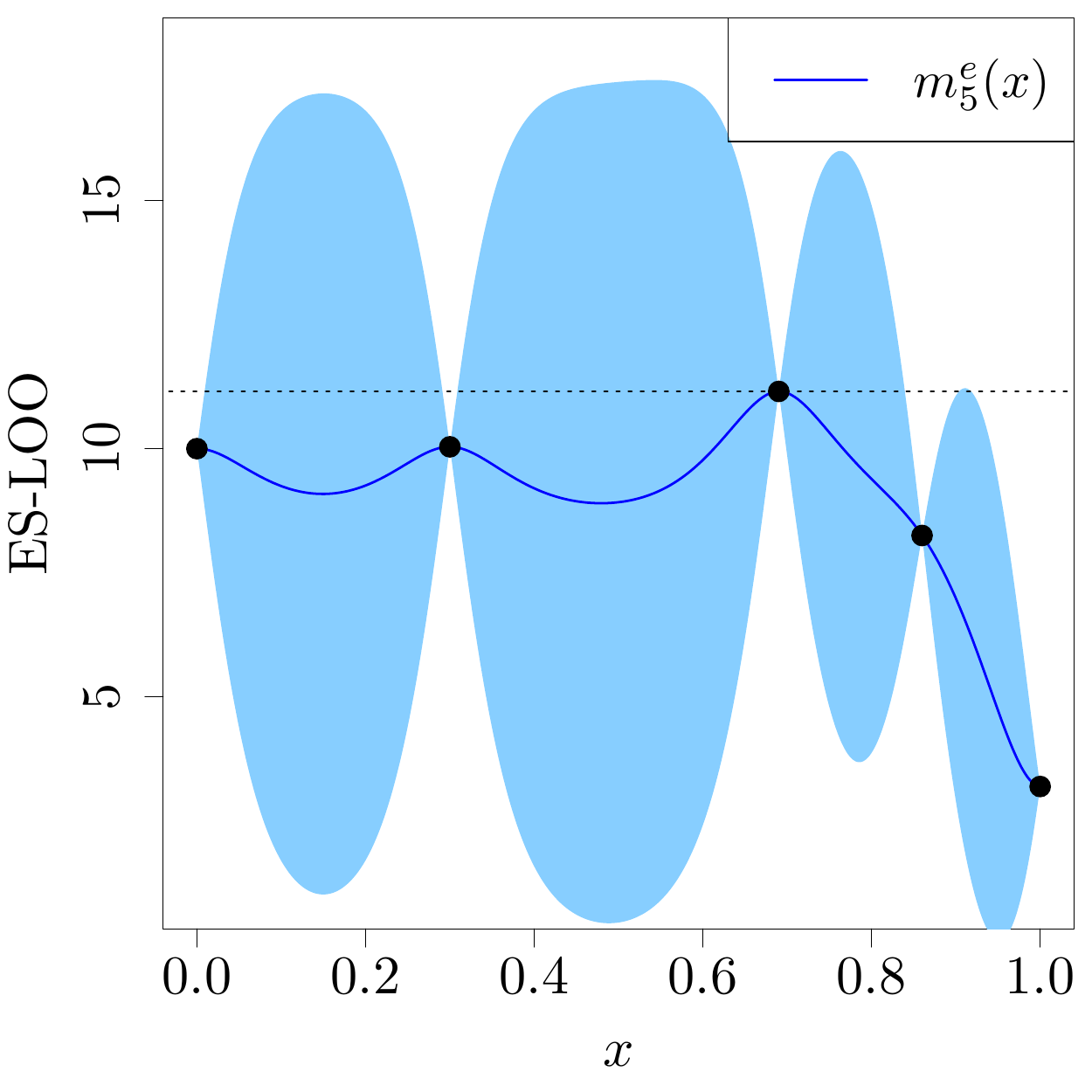}
	\includegraphics[width=0.49\textwidth]{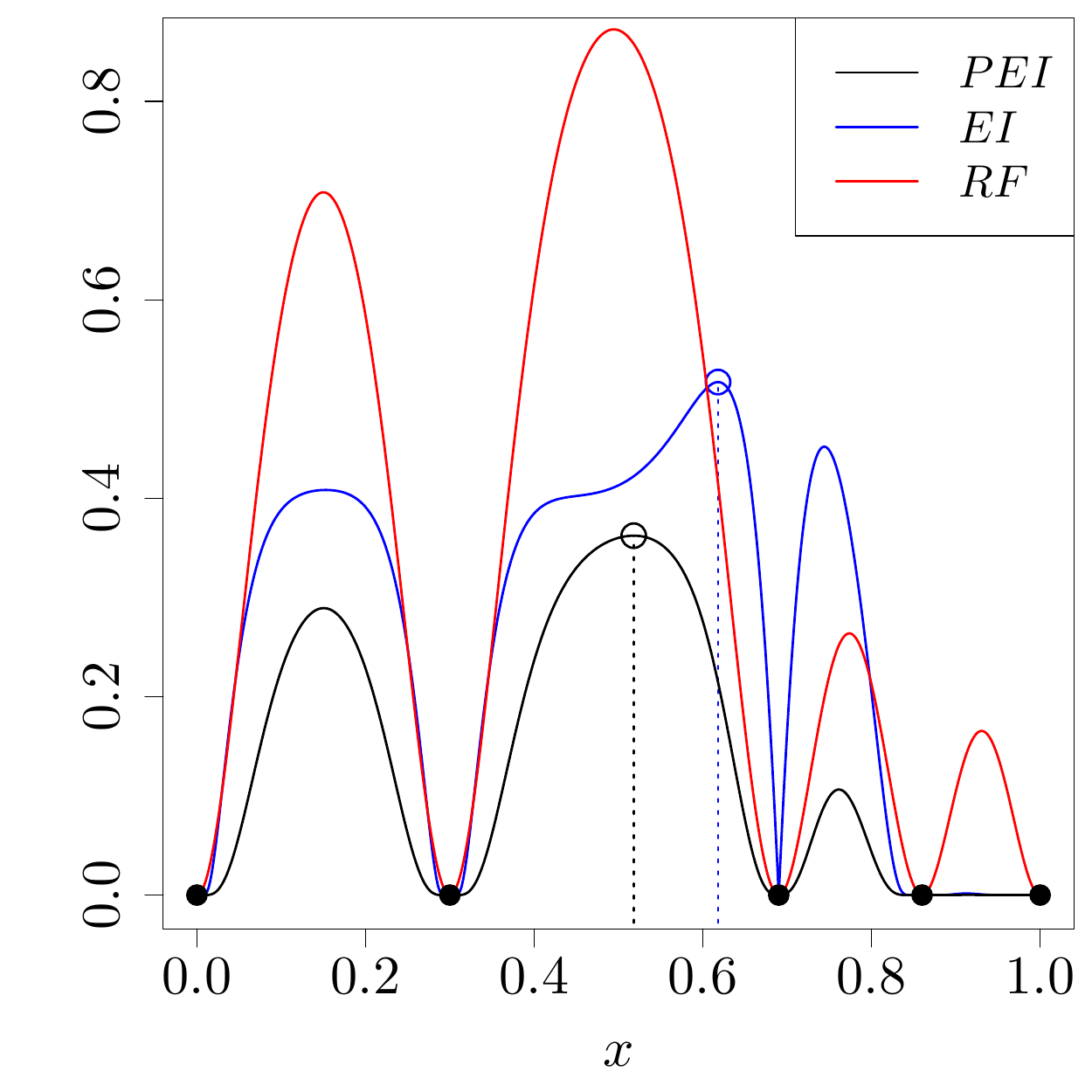}
	\caption{Left: A GP is fitted to five expected squared LOO errors. Right: Pseudo expected improvement (black) is obtained by multiplying expected improvement (blue) by the influence functions (red). The next sample point is where the PEI is maximum (black circle). The maximum of EI is shown by the blue circle. The PEI criterion is more explorative than EI.}
	\label{PEI_EI_IF}
\end{figure}

In this paper, \emph{pseudo expected improvement (PEI)} \cite{zhan2017} is considered as the selection criterion which has a better exploration property than EI. Pseudo expected improvement is obtained by multiplying EI by a repulsion function (RF): $PEI(\bx) =  EI(\bx) RF(\bx)$. The repulsion function is defined as   
\begin{equation}
	RF(\bx; \mathbf{X}_n) = \prod_{i = 1}^{n} \left[ 1 - \Cor \left(Z_n^e(\bx), Z_n^e(\bx_i) \right) \right] ,
	\label{RF} 
\end{equation}
where $\Cor (\cdot, \cdot)$ is the correlation function of $Z_n^e(\cdot)$. $ RF(\bx)$ is a measure of the (canonical) distance between $\bx$ and the design points. It is always between zero and one and zero at the data points because $\Cor \left(Z_n^e(\bx_i), Z_n^e(\bx_i) \right) = 1~, \forall \, \bx_i \in \mathbf{X}_n$. Multiplying EI by the repulsion function improves its exploration property as is shown in \ref{PEI_EI_IF}. The picture on the left illustrates a GP fitted on five data points and the corresponding repulsion function (red), EI (blue) and PEI (black) are visualised on the right picture. It can be seen that the maximum of EI is biased towards the design point whose ES-LOO error is maximum.

The correlation function in \ref{RF} depends on length-scales $\boldsymbol{\theta}^e = \lbrack \theta_1^e, \ldots, \theta_d^e \rbrack^\top$ that are estimated from $\lbrace \mathbf{X}_n, \mathbf{y}_n^e  \rbrace$. It is important that $\theta_i^e$s do not take very ``small" values to circumvent clustering. The reason is that when they come near zero, $\Cor \left(Z_n^e(\bx), Z_n^e(\bx_i) \right)$ tends to zero and $RF(\bx)$ is (almost) one everywhere meaning that it has no influence on EI. Besides, the maximum of EI is located in a shrinking neighbourhood of the current best point when $\boldsymbol{\theta}^e$ is small \cite{mohammadi2016}. Therefore, a lower bound has to be considered for $\boldsymbol{\theta}^e$. In a normalised input space, i.e. $\mathcal{D} = [0, 1]^d$, we define this lower bound to be 
\begin{equation}
	\theta_{lb}^e = \sqrt{-0.5/ \ln(10^{-8})} \, ,
	\label{theta_lb}
\end{equation} 
for each dimension. It is obtained by setting the minimum correlation equal to $10^{-8}$ for the squared exponential correlation function defined as
\begin{equation}
	k_0(x, x^\prime) = \exp \left(- \frac{|x - x^\prime|^2}{2\theta^2} \right).
	\label{SE_kernel}
\end{equation}
In the above equation, the minimum correlation between $x$ and $x^\prime$ happens when $|x - x^\prime| = 1$ which is the maximum distance between the two points in the normalised input space. 

\begin{figure}[htpb] 
	\centering
	\includegraphics[width=0.49\textwidth]{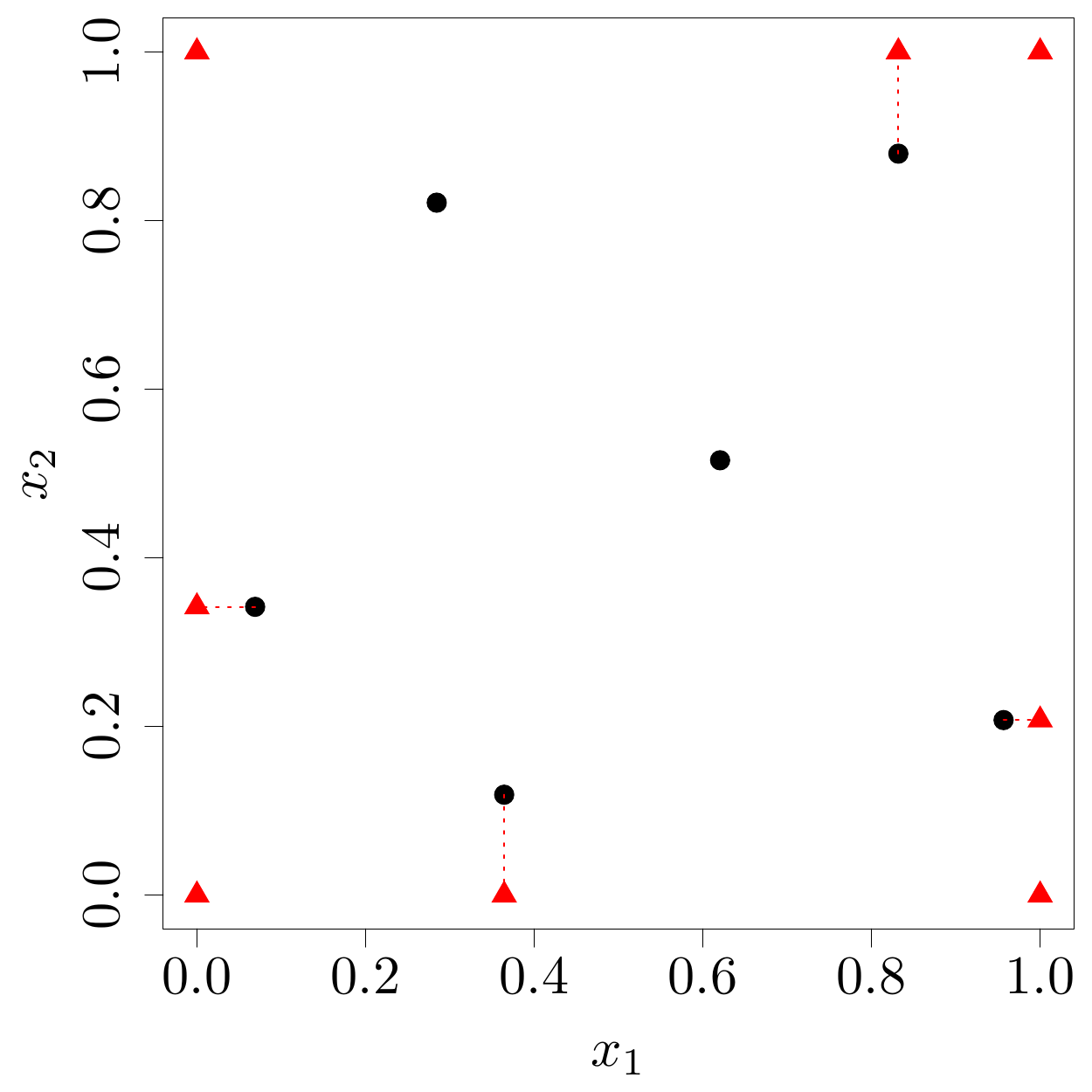}
	\caption{Initial design (black points) and pseudo points (red triangles). They are closest point on each face of the input space to the initial design and also at the corners of the region. The repulsion function is updated by pseudo points at almost no cost.}
	\label{pseud_point}
\end{figure}

A common issue in adaptive sampling strategies is that many design points lie along the boundaries of the input space where the predictive variance is large. Such samples might be nonoptimal if the true model is not feasible on the boundaries \cite{gramacy2009}. However, this problem can be alleviated in our approach by introducing ``pseudo points" at appropriate locations on the boundaries. Pseudo points, denoted by $\mathbf{X}_p$, are used to update the repulsion function, i.e. $RF(\bx; \mathbf{X}_n \cup \mathbf{X}_p)$, but the true function is not evaluated there due to computational cost. The following locations are considered as the pseudo points in our algorithm
\begin{enumerate}
	\item corners of the input space,
	\item closest point on each face of the (rectangular) bounding input region to the initial design.
\end{enumerate}
A 2-dimensional example is illustrated in \ref{pseud_point} that shows the location of pseudo points (red triangles) and six initial design (black points).  Finally, \ref{algorithm} summarises the steps of the proposed adaptive sampling method. 
\begin{algorithm}[htpb]
	\caption{Proposed sequential sampling approach}
	\label{algorithm}
	\begin{algorithmic}[1]
		\STATE Create an initial design: $\mathbf{X}_n = \left\{\textbf{x}_1, \dots , \bx_n \right\}$
		\STATE Evaluate $f$ at $\mathbf{X}_n$: $\mathbf{y}_n = f(\mathbf{X}_n)$
		\STATE Fit the GP $Z_n(\bx)$ to $\left\{\mathbf{X}_n, \mathbf{y}_n \right\}$
		\WHILE{\NOT stop}
		\FOR{$i=1$ to $n$}
		\STATE  Calculate $\ES(\bx_i)$  (\ref{exp_squar_LOO})
		\ENDFOR
		\STATE Set $\mathbf{y}_n^e = \left(\ES(\bx_1), \ldots, \ES(\bx_n) \right)^\top$
		\STATE Set the lower bound for $\boldsymbol{\theta}^e$ (\ref{theta_lb})
		\STATE Fit the GP $Z_n^e(\bx)$ to $\left\{\mathbf{X}_n, \mathbf{y}_n^e \right\}$
		\STATE Create pseudo points $\mathbf{X}_p$ (see \ref{pseud_point})
		\STATE $\bx_{n+1} ~\gets ~ \underset{\bx \in \mathcal{D}}{\arg\!\max}~ PEI(\bx) = EI(\bx) RF(\bx; \mathbf{X}_n \cup \mathbf{X}_p)$  
		\STATE Set $\mathbf{X}_n = \mathbf{X}_n \cup \left\{\bx_{n+1} \right\}$ 
		\STATE Evaluate $f$ at $\bx_{n+1}$ and $y_{n+1} ~\gets ~ f(\bx_{n+1})$
		\STATE Set $\mathbf{y}_n = \mathbf{y}_n \cup \left\{y_{n+1} \right\}$
		\STATE Update $Z_n(\bx)$ using $\left( \bx_{n+1}, y_{n+1} \right)$
		\STATE  $n \gets n+1$
		\ENDWHILE
	\end{algorithmic}
\end{algorithm}
\subsection{Extension to batch mode}
\label{sec:batch_sampling}
\begin{figure}[htpb] 
	\includegraphics[width=0.325\textwidth]{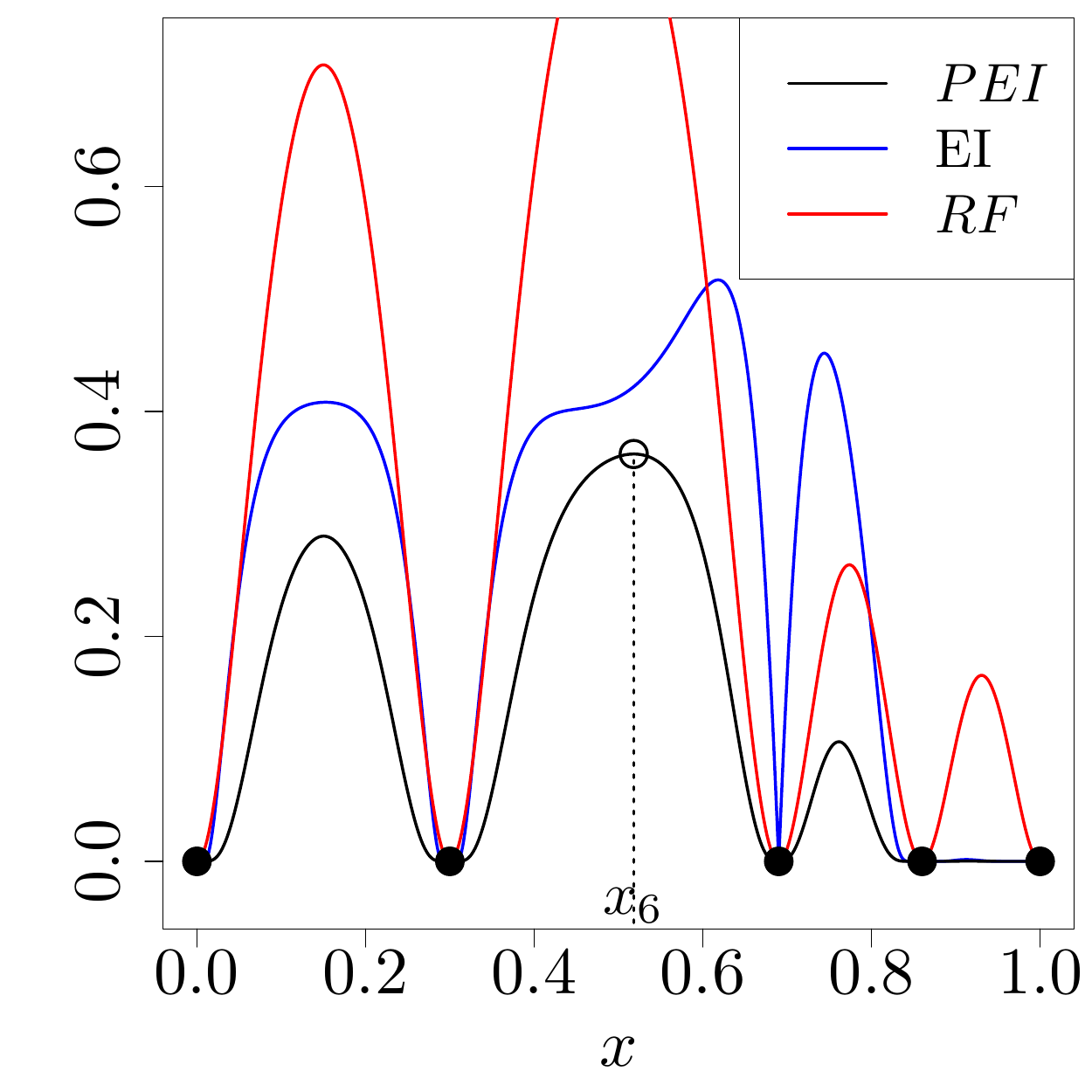}
	\includegraphics[width=0.325\textwidth]{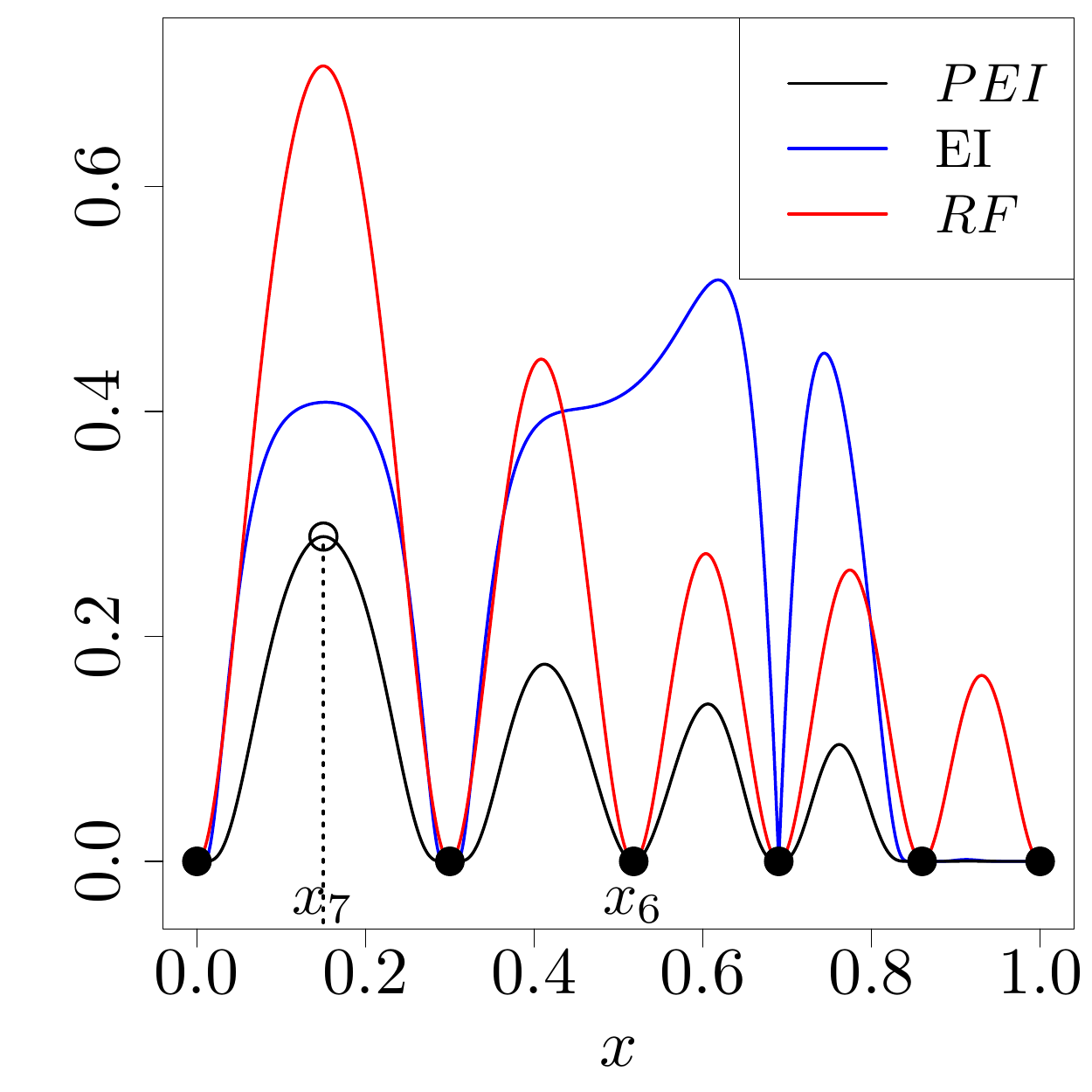}
	\includegraphics[width=0.325\textwidth]{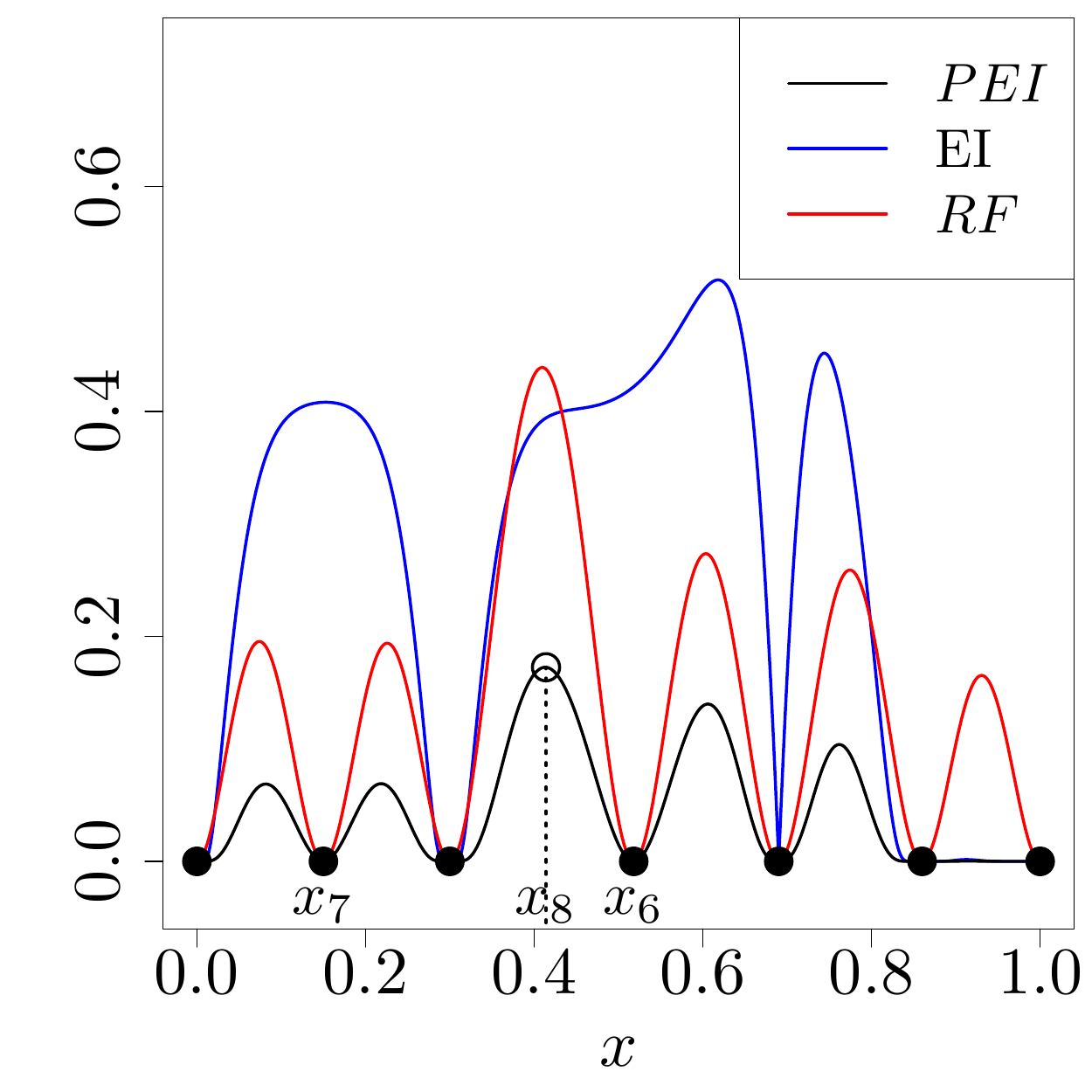}
	\caption{A batch of $q = 3$ points ($x_6, x_7$ and $x_8$) are selected thanks to the repulsion function. Left: there are five initial samples and the first query point, $x_6$, is chosen where PEI is maximum. Middle: the repulsion function is updated as $RF(x ; x_{1:5} \cup x_{6})$ which updates PEI accordingly without evaluating $f$ at $x_6$. The second query point, $x_7$, is the location of maximum of updated PEI. Right: PEI is updated using $x_7$ and the third sample, $x_8$, is selected as explained.}
	\label{batch_sampling}
\end{figure}

When parallel computing is available, it is often better to evaluate the expensive function $f$ at a set of inputs rather than a single point since it saves the user time. In batch sampling, $q > 1$ locations are chosen for evaluation at each iteration. Note that the computation time of running the simulator on $q$ parallel cores is the same as a single run. The PEI criterion can be employed in a batch mode thanks to the repulsion function. We now show how to choose $q$ points $\bx_{n+1}, \ldots, \bx_{n+q}$ in a single iteration. The first point $\bx_{n+1}$ is obtained by maximising the PEI criterion. Then, the repulsion function is updated by $\bx_{n+1}$ 
\begin{equation}
	RF(\bx ; \mathbf{X}_n \cup \bx_{n+1}) = \prod_{i = 1}^{n+1}  \left[  1 - \Cor \left(Z_n^e(\bx), Z_n^e(\bx_i) \right) \right] ,
	\label{IF_update}
\end{equation}
which updates PEI without evaluating $f$ at $\bx_{n+1}$. The second location $\bx_{n+2}$ is selected where the updated PEI is maximum. We repeat this procedure until the last point  $\bx_{n+q}$ is chosen
\begin{align*}
	& RF \left(\bx ; \mathbf{X}_n \cup \bx_{n+1} \cup \ldots \cup \bx_{n+q-1} \right) = \prod_{i = 1}^{n+q-1}  \left[  1 - \Cor \left(Z_n^e(\bx), Z_n^e(\bx_i) \right) \right] \\
	& \bx_{n+q} = \underset{\bx \in \mathcal{D}}{\arg \max} \, EI(\bx) \, RF \left(\bx ; \mathbf{X}_n \cup \bx_{n+1} \cup \ldots \cup \bx_{n+q-1} \right) . 
\end{align*}
\ref{batch_sampling} shows our adaptive sampling method in batch mode where $q = 3$ locations ($x_6, x_7$ and $x_8$) are picked in one iteration. The first new sample $x_6$ is chosen where the PEI criterion is maximum (left). Then, the repulsion function is updated by $x_6$. This will update PEI and its maximum allows us to find $x_7$ without evaluating $f$ at $x_6$ (middle). Again, we update the repulsion function using $x_7$ and maximise the updated PEI to obtain the sample site $x_8$ (right).
\section{Numerical experiments} 
\label{sec:experiments}
\begin{figure}[htpb] 
	\centering
	\includegraphics[width=0.47\textwidth]{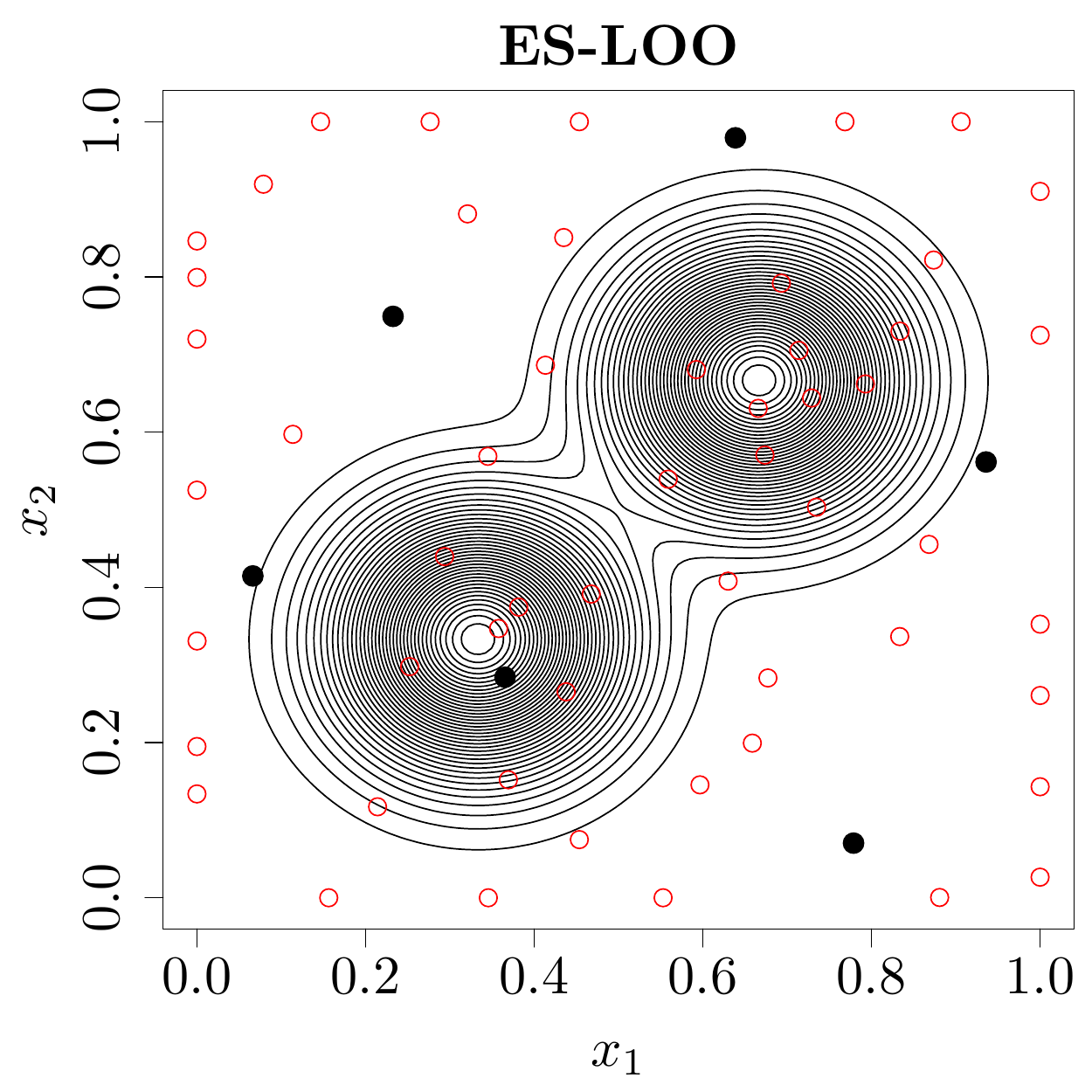}
	\includegraphics[width=0.47\textwidth]{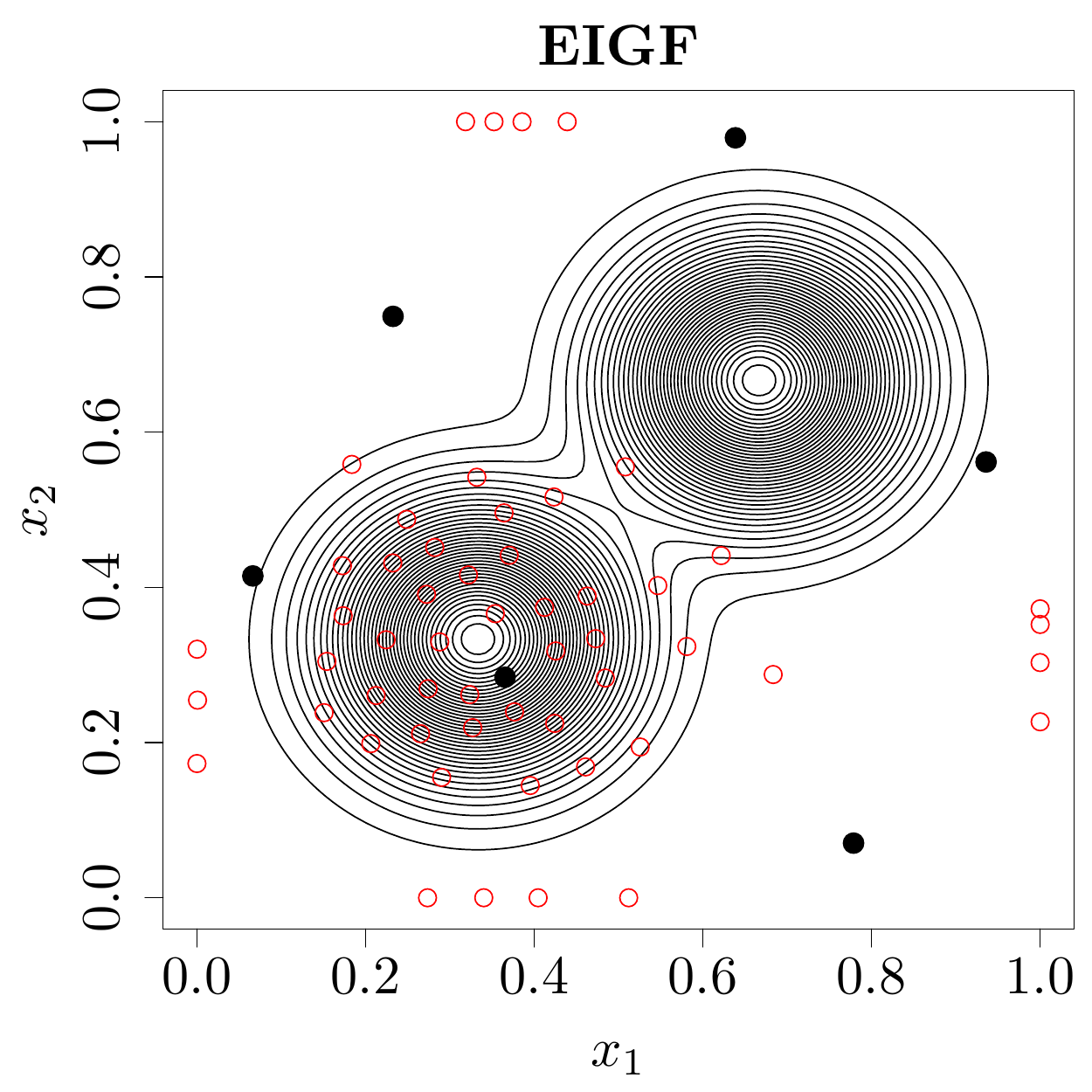}  \\
	\vspace{0.5cm}
	\includegraphics[width=0.47\textwidth]{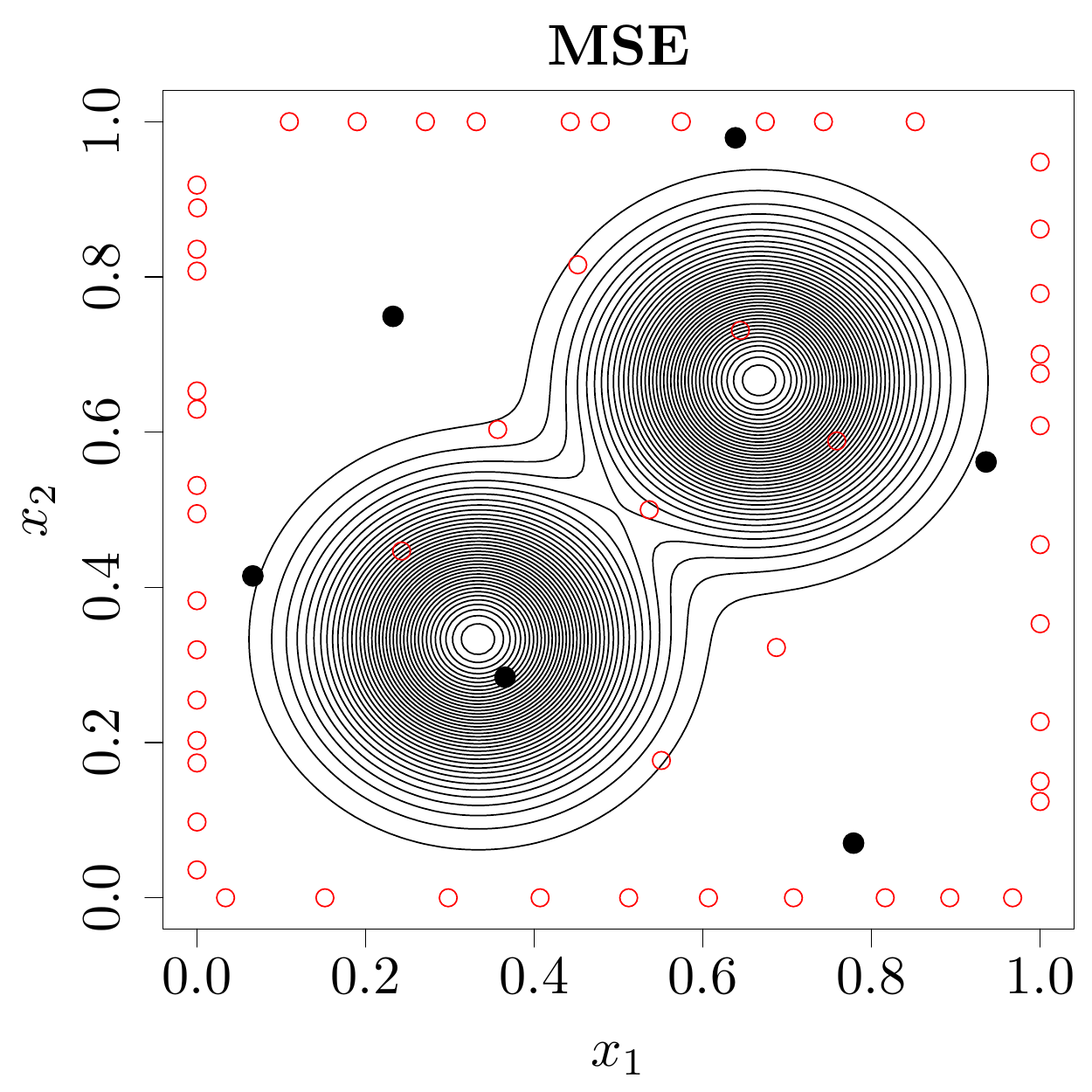} 
	\includegraphics[width=0.47\textwidth]{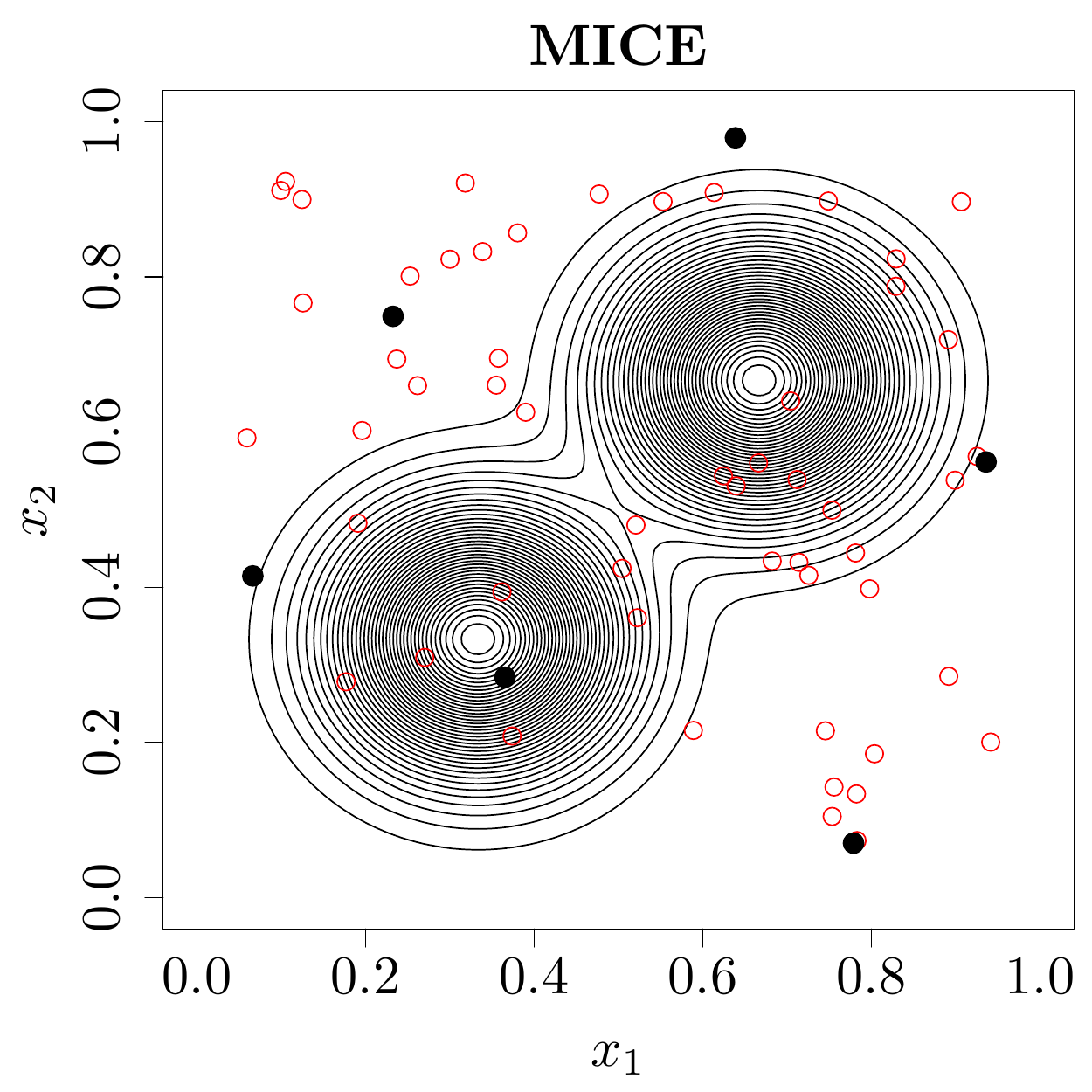} 
	\caption{A typical sampling behaviour of four adaptive sampling methods: ES-LOO, EIGF, MSE and MICE.  The red circles represent adaptive samples that are added to the six initial design (black dots). ES-LOO tends to fill  the space with focusing on regions where the function values change rapidly. EIGF mainly exploits the basin of attraction of an optima while other areas are unexplored. MSE samples more points on the boundaries where the prediction uncertainty is large. There is not any special pattern in the sampling behaviour of MICE, however, it avoids to put points on the boundaries.}
	\label{2D_adaptive_approches}
\end{figure}

Experimental results are presented and discussed in this section. Four analytic test functions and two real-world problems are considered as the ``true" function to assess the efficiency of our proposed sampling method. The results are compared with one-shot Latin hypercube sampling (LHS) and three adaptive approaches, namely MSE, EIGF and MICE. A typical characteristic of these adaptive sampling approaches is exhibited in \ref{2D_adaptive_approches} where the black dots are the initial design (identical in all pictures) and the red circles represent the adaptive samples. The true function has two spikes as it is a sum of two Gaussian functions centred at $\left(1/3, 1/3\right)^\top$ and $\left(2/3, 2/3\right)^\top$. As can be seen, our proposed method fills space uniformly with a focus on areas where the true function has nonlinear behaviour. The efficiency of our algorithm is due to two parts together. Firstly, the ES-LOO measure that reflects the actual model error and indicates promising regions for future evaluations. Secondly, incorporating the repulsion function in our EI-based selection criterion that prevents points from clustering. Without the repulsion trick, as is the case in \ref{franke_EI}, new experimental designs tend to pile up at points where the ES-LOO errors are large. However, further investigation is required to identify the exact contribution of each part on the algorithm. In our method the boundary issue, i.e. taking many points on the boundaries, is mitigated by introducing pseudo points at locations shown in \ref{pseud_point}. 

EIGF tends toward local exploitation and does not explore the input space; it gets stuck in an optimum. This can be explained according to the EIGF formula in \ref{EIGF_criterion} where local exploitation is carried out by the first term, $\left(m_n(\bx) - f(\bx_i^*) \right)^2$. It represents the difference between the GP prediction and response at the nearest sample location $\bx_i^*$ and increases in regions where a drastic response change takes place. As a result, the first term can be interpreted as a measure of gradient. In \ref{2D_adaptive_approches}, there is a design point close to the centre of the spike $\left(1/3, 1/3\right)^\top$ at which the function reaches its maximum. Since the function varies ``significantly" there, the EIGF criterion is large (due to its first term) around $\left(1/3, 1/3\right)^\top$ where most samples are taken. Note that the points sampled on the boundaries are due to increase in the second term of the EIGF expression, i.e. $s_n^2(\bx)$. MSE samples most points on the boundaries where the prediction uncertainty is large. It is possible that MSE leads to a space-filling design as is the case of the Franke's function, see below. No special trend can be found on the performance of the MICE algorithm. However, it avoids sampling around the boundaries and all new points are taken in the interior region. We describe our implementation of ES-LOO below.
\subsection{Experimental setup}
\label{sec:setup}
The prediction accuracy is assessed by the \emph{root mean squared error (RMSE)} criterion. Given the test set $\lbrace \left(\bx_t, f(\bx_t) \right) \rbrace_{t = 1}^{t = N}$, RMSE is defined as
\begin{equation}
	RMSE = \sqrt{\frac{\sum_{t = 1}^{N} \left( m_n(\bx_t) - f(\bx_t) \right)^2 }{N}} \, ,
\end{equation}
which measures the distance between the emulator, $m_n(\bx_t)$, and the true function, $f(\bx_t)$. In our experiments, $N = 3000$ and the test points are selected uniformly across the input space. A total budget equal to $30 d$ is considered for each experiment. The initial space-filling DoE is of size $3 d$ and is obtained by the \texttt{maximinESE\_LHS} function implemented in the R package \emph{DiceDesign} \cite{DiceDesign}. There are ten different initial DoEs for every function and we assess the prediction performance of each method using all ten sets. The R package \emph{DiceKriging} \cite{roustant2012} is employed to construct GP models. The covariance kernel for modelling the true function and $\ES(\bx)$ is Mat\'ern with $\nu = 3/2$. Since the ES-LOO error is a positive quantity, the GP emulator is fitted to its natural logarithm. The closest point on each face of the input space to the initial design and the corners of the input region are considered as the pseudo points. 

The optimisation of the PEI function (and other selection criteria) is conducted by the differential evolution (DE) algorithm \cite{storn1997} implemented in the R package \emph{DEoptim} \cite{mullen2011}. This algorithm is considered by \cite{zhan2017} to optimise acquisition functions used in the Bayesian optimisation paradigm. DE is a stochastic global search algorithm for continuous problems and belongs to the family of population-based evolutionary methods. It can tackle nondifferentiable, multimodal functions and has a small number of parameters to tune. The progress toward better solutions in the search space is made by applying mutation, crossover and selection operators to the population of candidate solutions. In this paper, the population size is equal to $10d$. The other control parameters (such as the step size and crossover probability) are set to their default values, see the \emph{DEoptim} package.
\subsection{Test functions}
\label{sec:test_funs}
\begin{figure}[htpb] 
	\centering
	\includegraphics[width=0.47\textwidth]{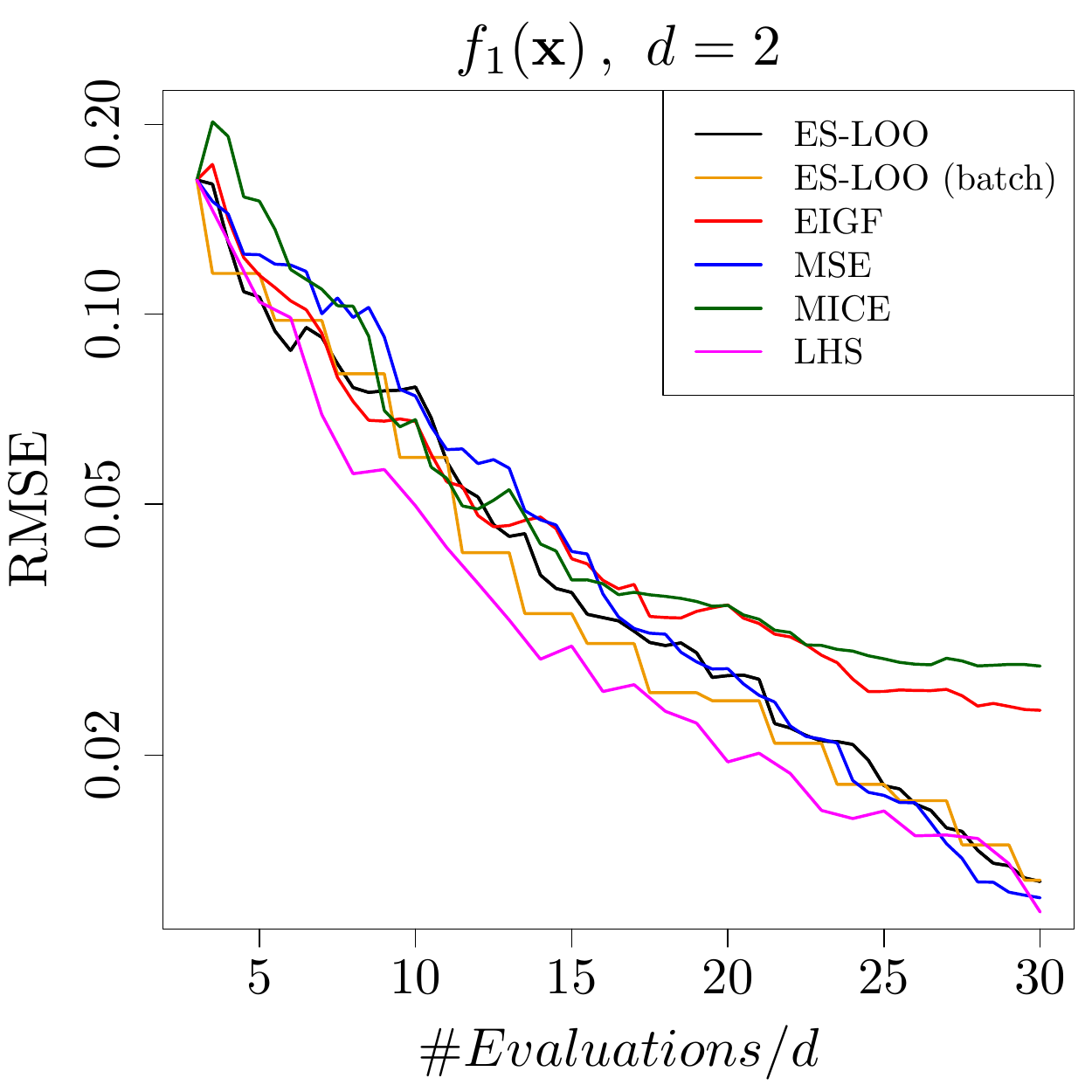}
	\includegraphics[width=0.47\textwidth]{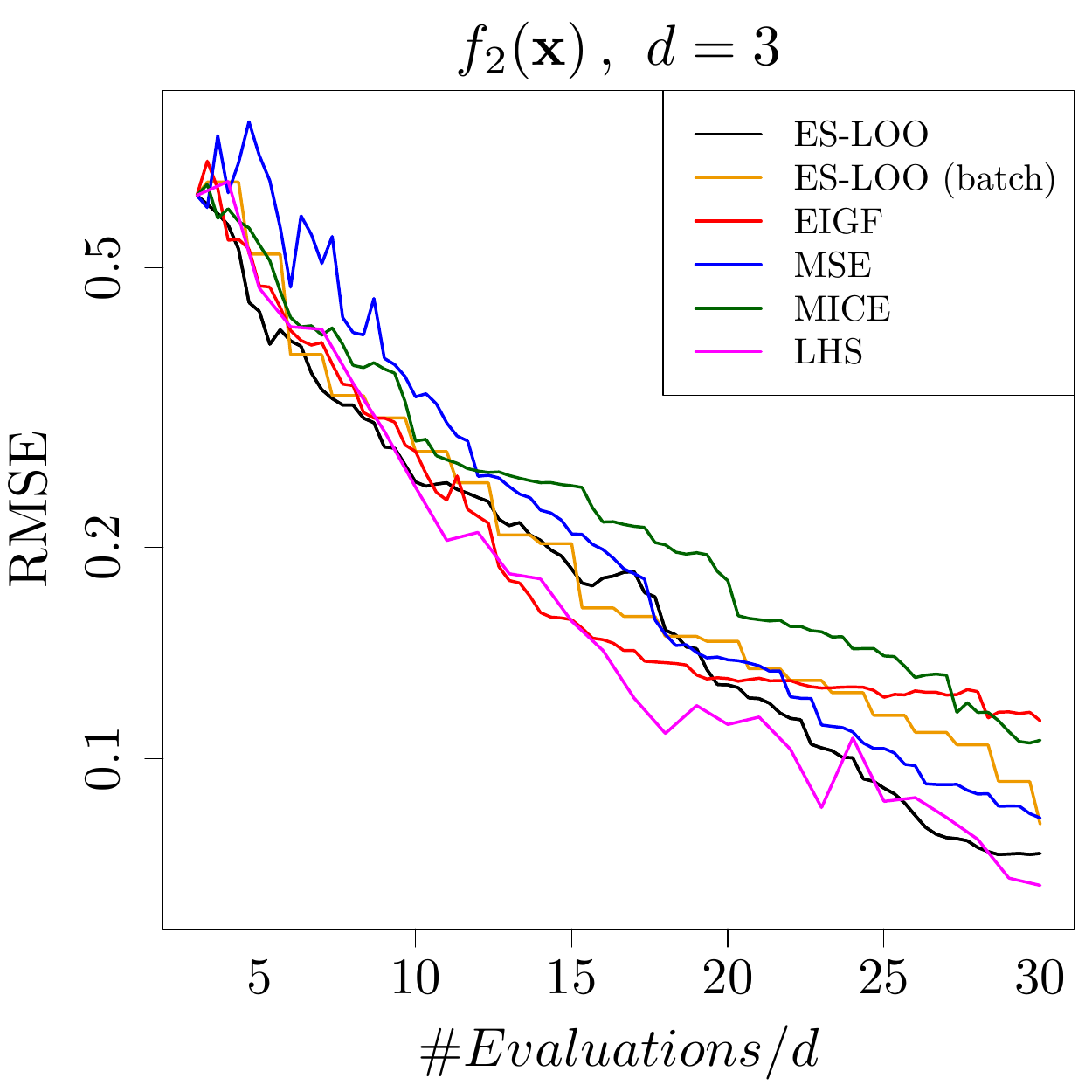}  \\
	\vspace{0.5cm}
	\includegraphics[width=0.47\textwidth]{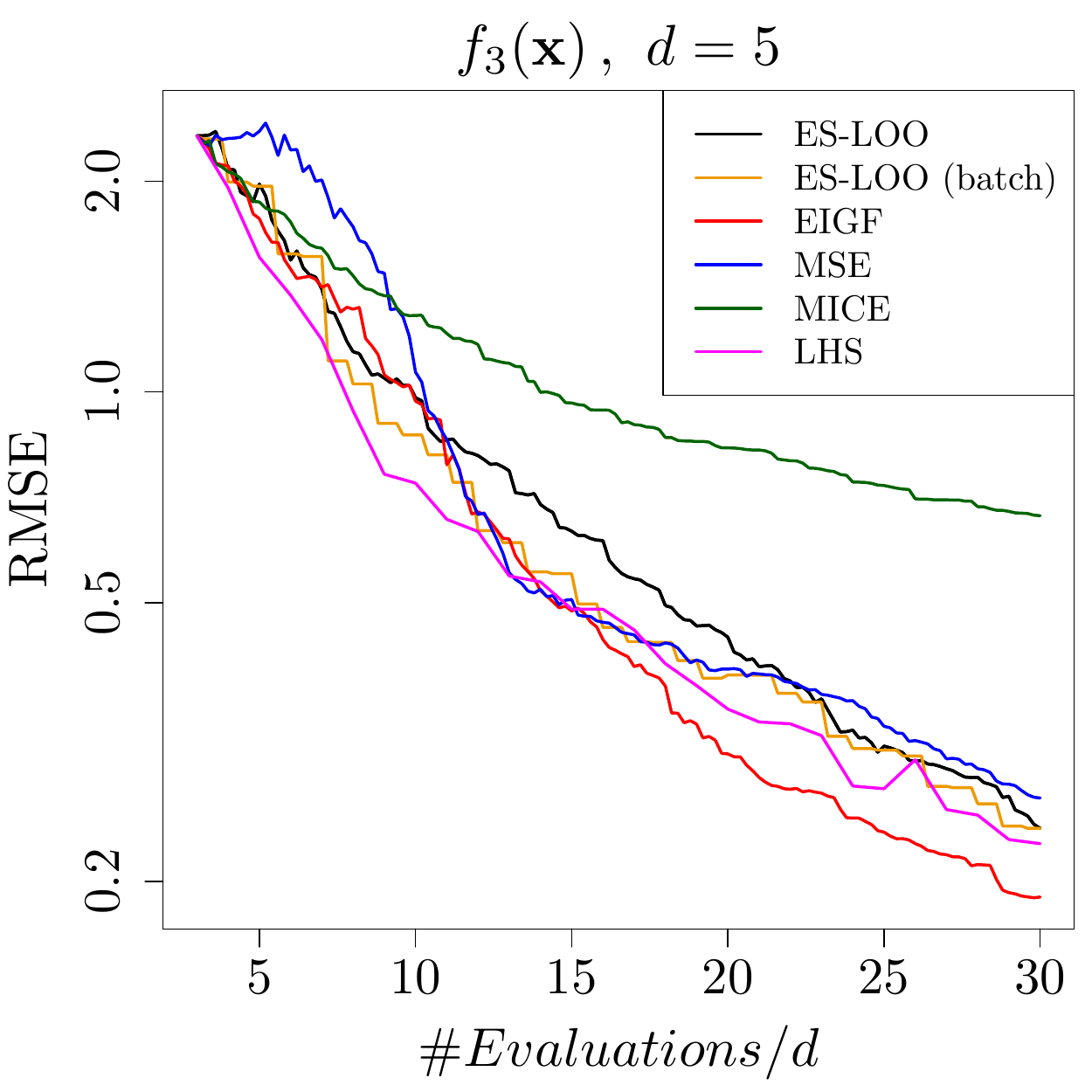} 
	\includegraphics[width=0.47\textwidth]{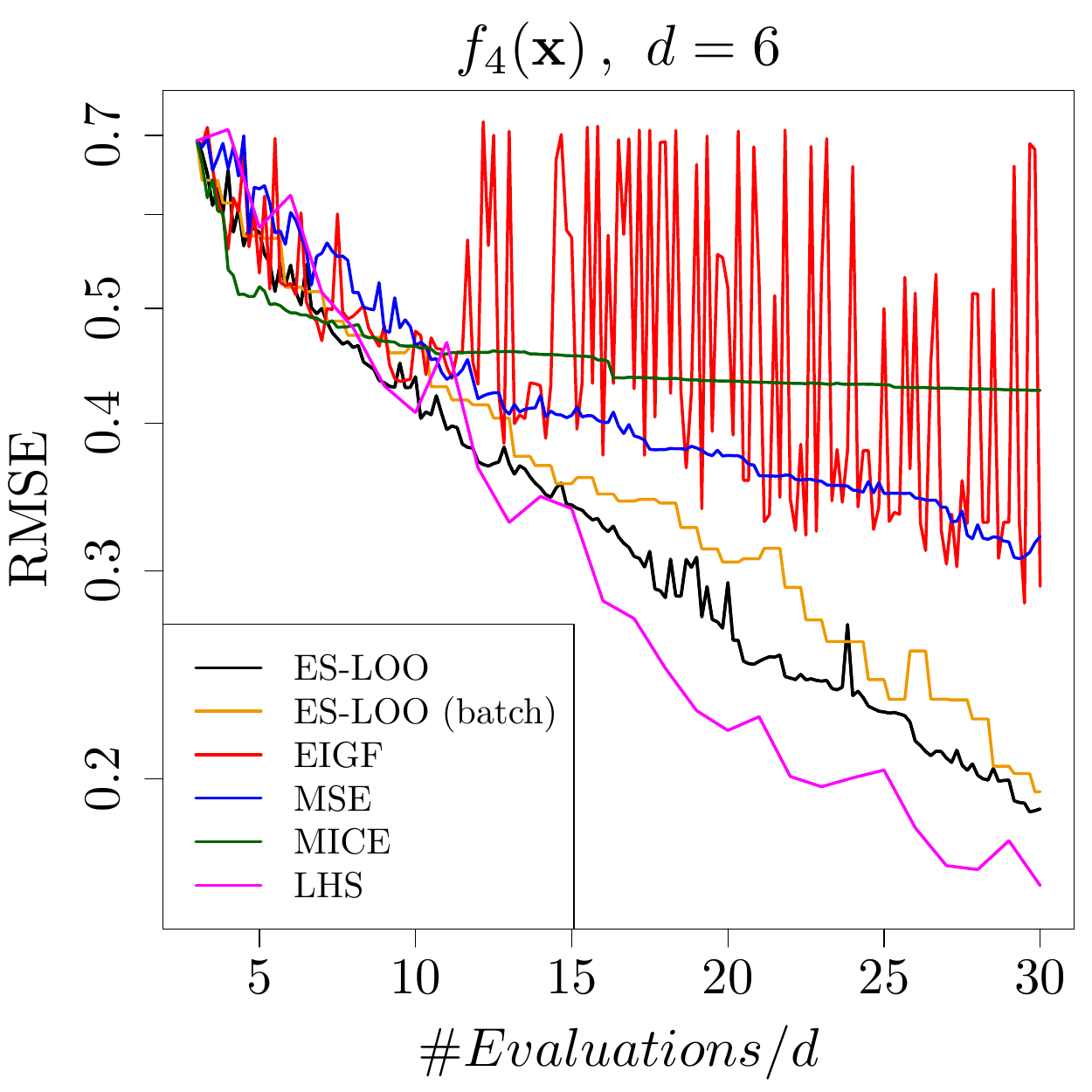} 
	\caption{The median of ten RMSEs of our proposed approach ES-LOO: sequential (black) and batch with $q = 4$ (orange), EIGF (red), MSE (blue) and MICE (green). Ten different initial DoEs of size $3d$ are considered for each function; every method produces ten predictions based on them. The median of RMSEs based on (one-shot) LHS design for sample sizes $3d$, $4d, \ldots, 30d$ is shown in magenta.  The total budget is $30d$. The $y$-axis is on logarithmic scale.}
	\label{result_test_fun}
\end{figure}

The four test functions are 
\begin{itemize}
	\item $f_1(\bx)$: Franke's function \cite{haaland2011}, $d = 2$
	\item $f_2(\bx)$: Hartman function \cite{jamil2013}, $d = 3$
	\item $f_3(\bx)$: Friedman function \cite{friedman1991}, $d = 5$
	\item $f_4(\bx)$: Gramacy \& Lee function \cite{gramacy2009}, $d = 6$
\end{itemize}   
and their analytic expression are given in \ref{toy_tests}. The four test functions are defined on $\left[0, 1\right]^d$. \ref{result_test_fun} illustrates a comparison between the prediction performance of our proposed method ES-LOO: sequential (black) and batch with $q = 4$ (orange), and three other sequential sampling approaches: MSE (blue), EIGF (red) and MICE (green). Each curve represents the median of ten RMSEs using ten different initial DoEs. The results are also compared with the one-shot space-filling design (magenta) for all sample sizes $3d$, $4d, \ldots, 30d$; the graph represents the median of ten RMSEs obtained by LHS. The $x$-axis shows the number of function evaluations divided by the problem dimension, $d$. The $y$-axis is on logarithmic scale. 

Generally, the prediction performance of our method is comparable to other adaptive approaches. In particular, it outperforms MSE and EIGF in approximating the Hartman ($f_2$) and Gramacy \& Lee ($f_4$) functions. As can be seen, the sequential (black) and batch (orange) ES-LOO have similar performances such that in both algorithms the RMSE is monotonically (almost linearly) decreasing in all test problems. However, this is not the case of other adaptive sampling approaches. The Hartmann function has four local minima and EIGF can stuck in one of them. The EIGF method has the best performance on the Friedman function ($f_3$) and has the lowest accuracy in approximating other functions, especially $f_4$. The reason is that the main response change of $f_4$ occurs near the right bound of the input space and EIGF puts more points there. The MICE algorithm shows a poor performance in predicting the test functions. However, it is the fastest algorithm as the criterion is based on a discrete representation. Other adaptive sampling methods could be implemented on similar discrete representation and they would then gain the speed up. It is observed that the points obtained by the MSE method fill the space almost uniformly in the Franke's function and thus has a similar performance to the LHS method. However, MSE favours sampling more points on the boundary of the input space of the (6-dimensional) $f_4$ function. LHS is the best sampling strategy to predict $f_1, f_2$ and $f_4$. However, the RMSE of ES-LOO approaches that of LHS when the number of function evaluations increases, say after $25\times d$ evaluations.

According to \ref{result_test_fun}, it seems that LHS should be preferred over adaptive sampling methods as it has a superior performance in most cases. However, this is not always the case as LHS performs poorly on the real-world problems presented in the next section. Besides, the main disadvantage of LHS (one-shot methods in general) is that we cannot stop them as soon as the emulator reaches an acceptable level of accuracy. In adaptive approaches, however, the costly sampling procedure can be halted at any time. For example, suppose that $f_2$ is the underlying function of a complex code and the LHS  sample size (as a rule of thumb) is: $10\times d = 30$. Let RMSE $= 0.5$ be the acceptable prediction accuracy. In this setting, the ES-LOO algorithm can be stopped at about $15$ runs which is half of the time required to execute the model based on the LHS design.
\subsection{Real-world problems}
\label{sec:real-world}
\begin{figure}[htpb] 
	\centering
	\includegraphics[width=0.47\textwidth]{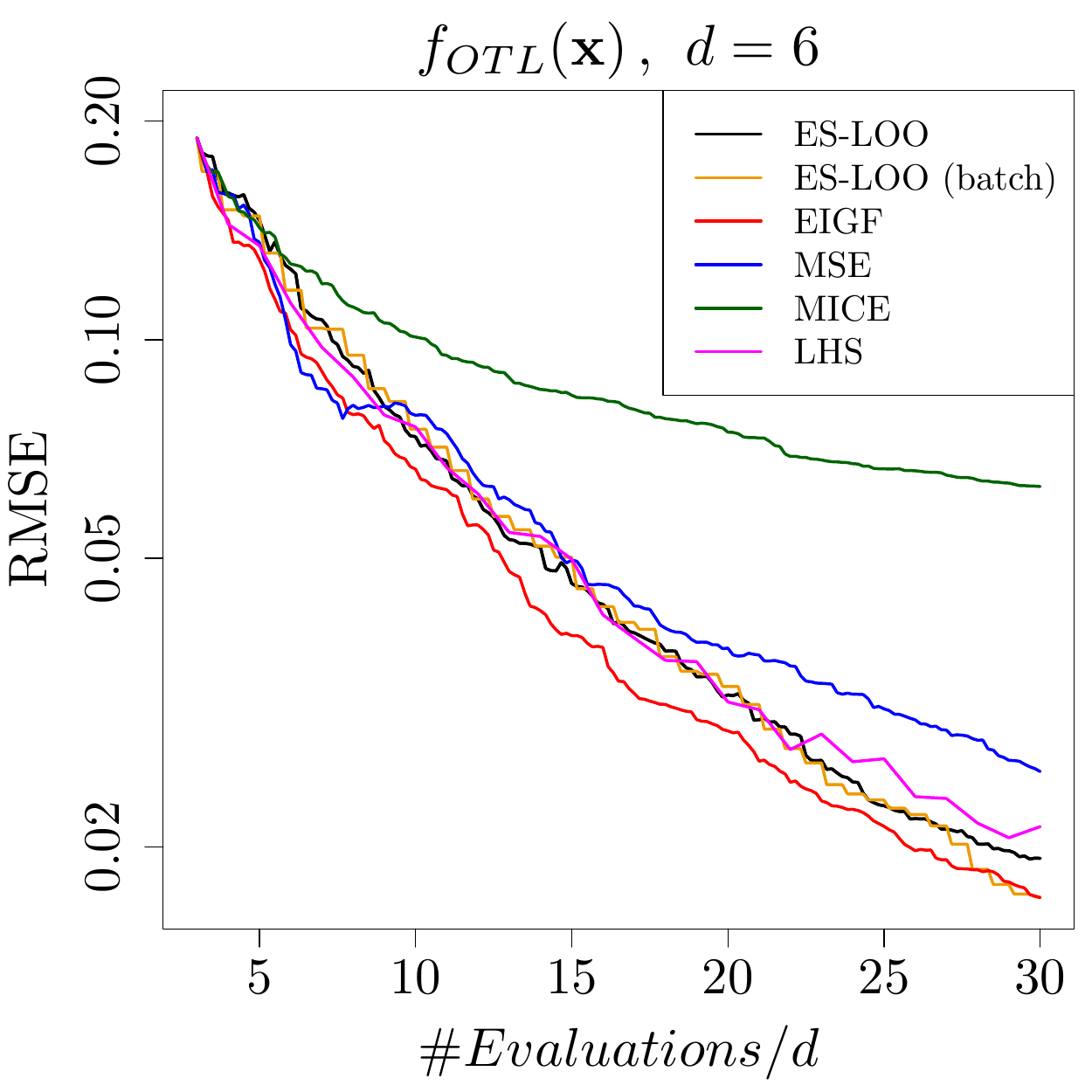}
	\vspace{0.5cm}
	\includegraphics[width=0.47\textwidth]{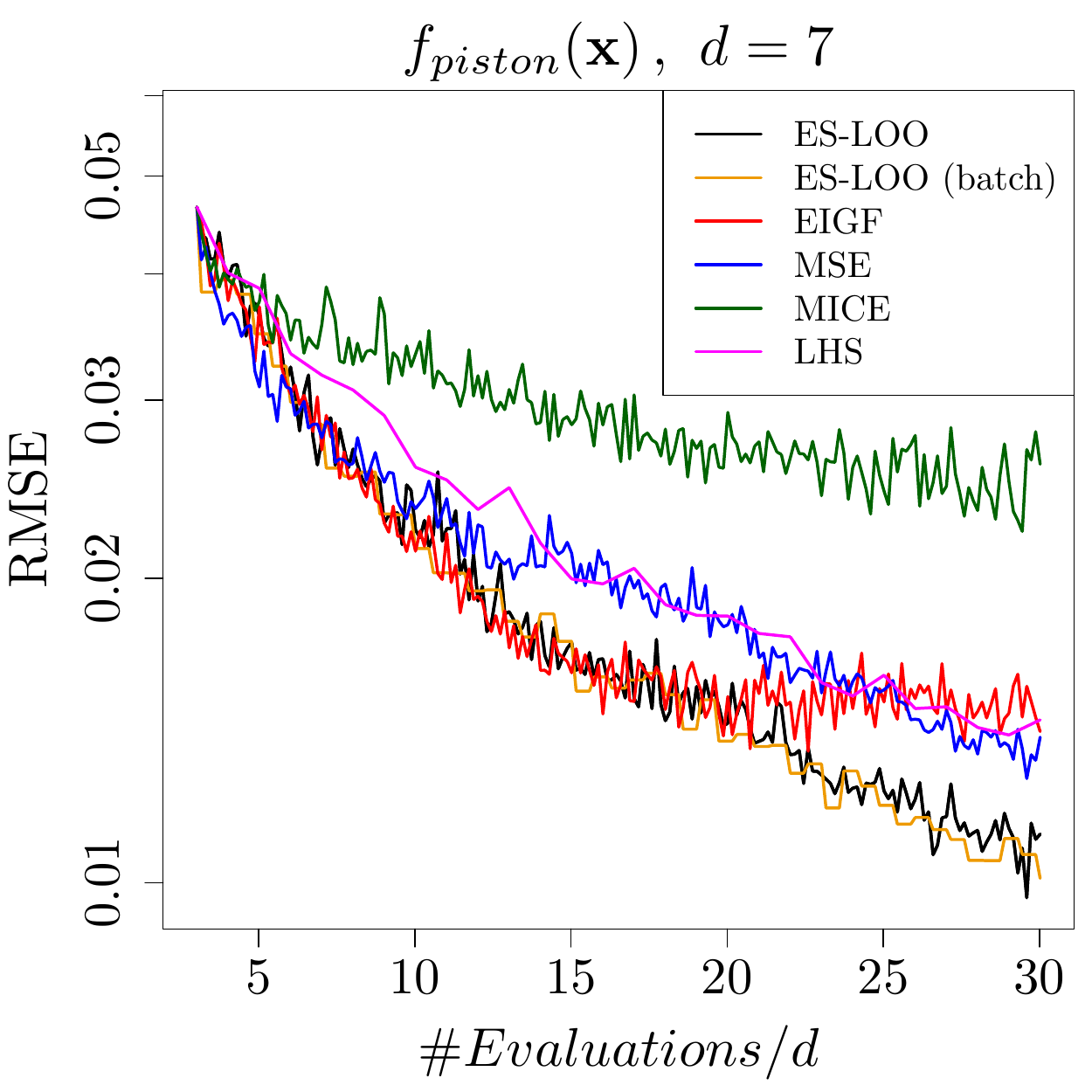} 
	\caption{The median of ten RMSEs of our proposed approach ES-LOO: sequential (black) and batch with $q = 4$ (orange), EIGF (red), MSE (blue) and MICE (green). Size of initial design $= 3d$ and total budget $= 30d$. The magenta line represents the median of RMSEs obtained by (one-shot) LHS design for sample sizes $3d$, $4d, \ldots, 30d$. The $y$-axis is on logarithmic scale.}
	\label{result_real-world}
\end{figure}

We also tested our method on two real-world problems which are the 6-dimensional \emph{output transformerless (OTL) circuit} and 7-dimensional \emph{piston simulation} functions \cite{benamri2007}. The former ($f_{OTL}$) returns the midpoint voltage of a transformerless circuit and the latter ($f_{piston}$) measures the cycle time that a piston takes to complete one cycle within a cylinder. The analytical expressions of $f_{OTL}$ and $f_{piston}$ and their design spaces are given in \ref{real_world_tests}. \ref{result_real-world} illustrates the results of comparing our proposed method with MSE, EIGF, MICE and LHS design in predicting the OTL circuit and piston simulation functions. The experimental setup is the same as explained in \ref{sec:setup}. The sequential (black) and batch (orange) ES-LOO have again similar performances and their RMSE criterion reduces steadily on both problems. 

As can be seen, ES-LOO is the best sampling approach for emulating the piston simulation function and performs quite well on the OTL problem. The MSE approach does not work well in both problems, especially after almost $10\times d$ function evaluations. Based on our experiments, MSE is not recommended when the dimensionality of the problem is larger than five. LHS has a similar behaviour to MSE and is not a good sampling approach on the OTL and piston simulation problems, contrary to our previous results demonstrated in \ref{result_test_fun}. While EIGF has the best performance on the OTL circuit problem, there is no improvement in the prediction accuracy of the emulator after $10\times d$ evaluations of the piston simulation function. MICE is the least successful algorithm in these two problems. 
\section{Conclusions}
\label{sec:conclusion}
This paper deals with the problem of extending an initial design sequentially for training GP models. This is an important issue in the context of emulating computationally expensive computer codes where the goal is to approximate the underlying function with a minimum number of evaluations.
An adaptive sampling scheme is presented based on the expected squared leave-one-out error (ES-LOO) which is used to identify ``good" locations for future evaluations. Since the value of ES-LOO is only known at the design points, another GP model is applied to approximate it at unobserved sites. Then, the pseudo expected improvement criterion is employed at each iteration to find the location of maximum of ES-LOO as the most promising point to improve the emulator. Pseudo expected improvement is obtained by multiplying the expected improvement criterion by a repulsion function. Once the new sample is chosen, it is added to the existing designs and the procedure is repeated until a stopping criterion is met. The proposed method can be easily promoted to a batch mode where at each iteration a set of input points is selected for evaluation. This can save the user time if parallel computing is available. Several test functions are used to test the capability of our method and the results are compared with other commonly used sampling techniques. The results show that our proposed adaptive sampling approach is promising. Ideally there would be some proof of the asymptotic convergence of the ES-LOO to zero error. Unfortunately, we have not been able to derive such proofs and they remain for further work. We think that following similar results from the convergence of EI in Bayesian optimisation, see e.g. \cite{vazquez2010, bull2011, ryzhov2016}, can help to prove the asymptotic properties of our method.

\begin{figure}[htpb] 
	\centering
	\includegraphics[width=0.47\textwidth]{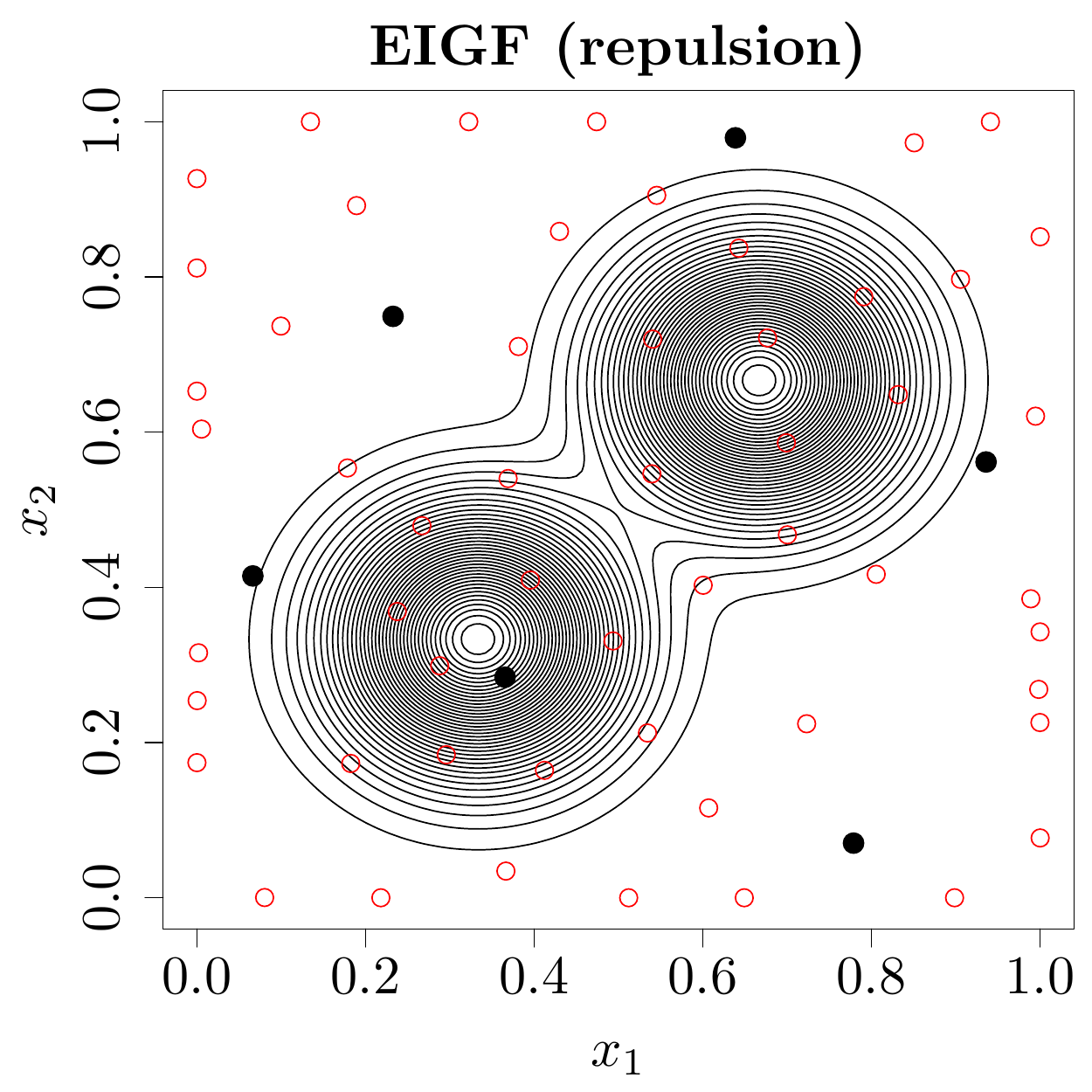}  
	\includegraphics[width=0.47\textwidth]{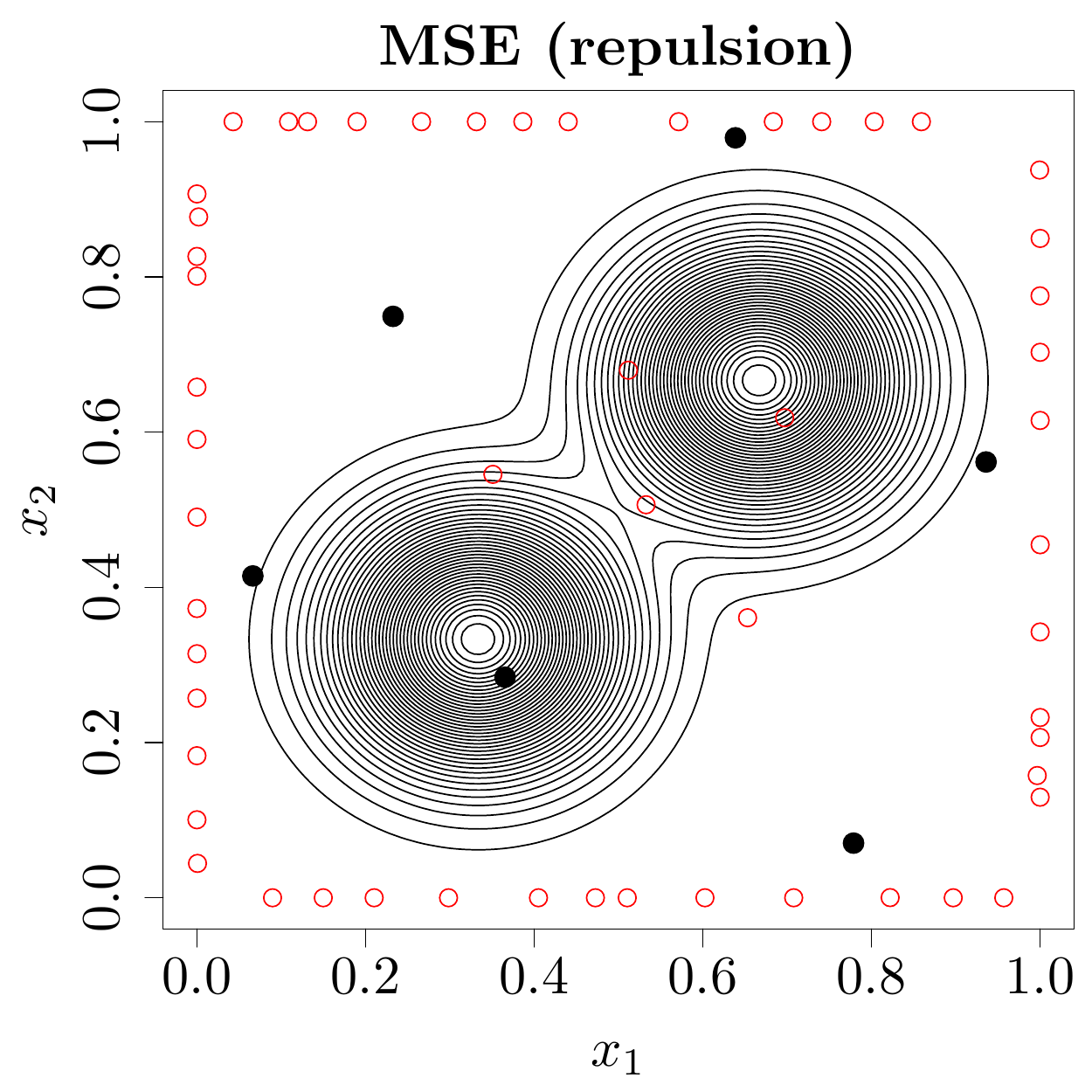}  \\
	\includegraphics[width=0.47\textwidth]{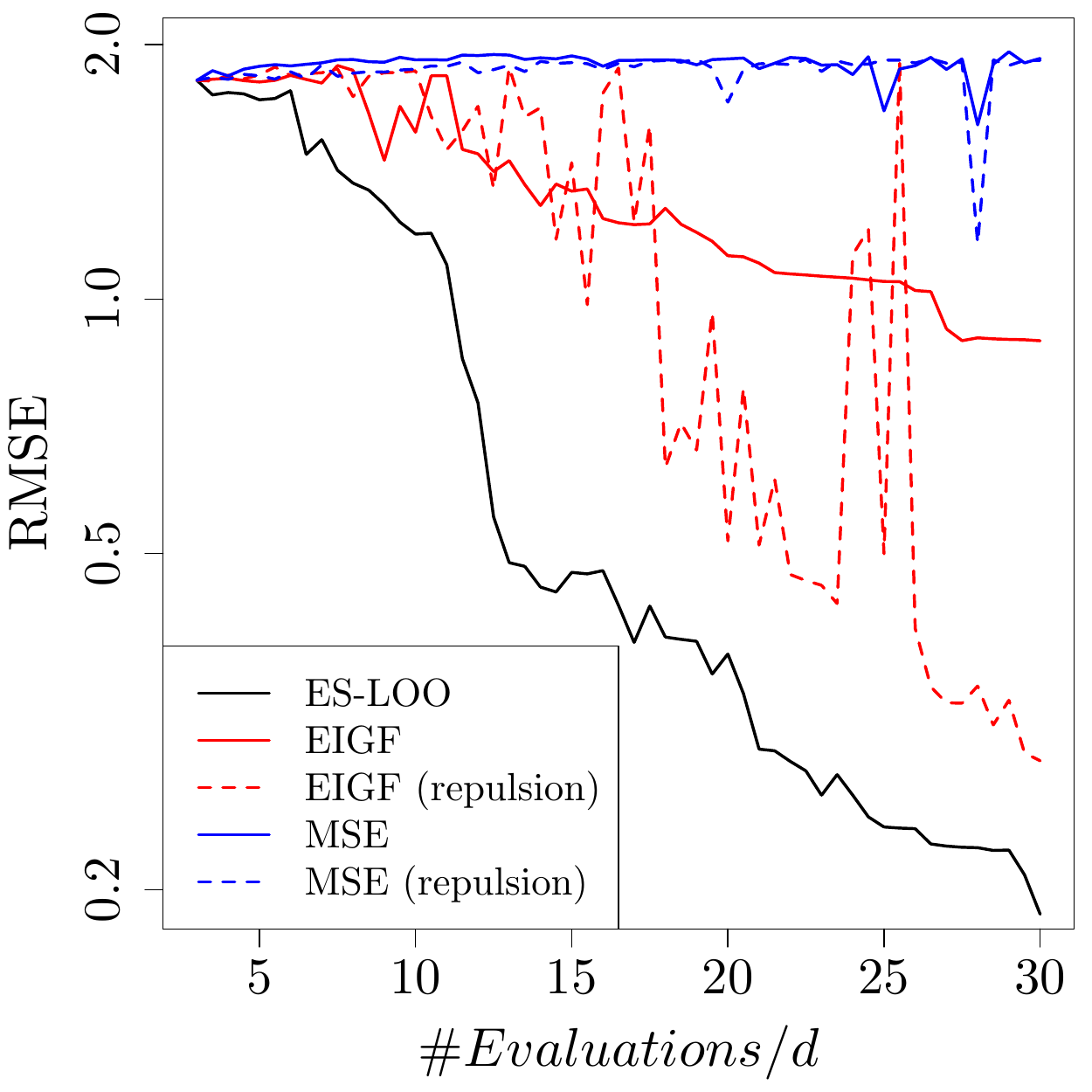}
	\caption{Top: Incorporating the repulsion function in the EIGF (left) and MSE (right) algorithms. While such modification promotes the diversity of points in EIGF, it intensifies the boundary issue in MSE. Bottom: The median of ten RMSEs associated with ES-LOO (black), EIGF (red), MSE (blue), modified EIGF (dashed red) and MSE (dashed blue) with the repulsion function.}
	\label{2D_adaptive_approches_repulsion}
\end{figure}

One can extend the idea of the repulsion function to other sequential sampling approaches. For example, \ref{2D_adaptive_approches_repulsion} illustrates the behaviour of EIGF and MSE modified by the repulsion trick. As can be seen, such modification promotes the diversity of samples in EIGF and improves the RMSE criterion. However, in the modified MSE algorithm the boundary issue is intensified as the repulsion function tends to push away the points. A possible future research direction is to investigate the inclusion of such repulsion-type functions in sampling strategies to improve their exploration property.
\section*{Acknowledgments}
The authors gratefully acknowledge the financial support of the EPSRC via grant EP/N014391/1 and of the Alan Turing Institute. The authors would like to thank the Isaac Newton Institute for Mathematical Sciences, Cambridge, for support and hospitality during the Uncertainty Quantification programme (UNQ) where work on this article was undertaken. We would like to thank the two anonymous reviewers for their suggestions and comments. We also warmly thank Joakim Beck for discussions on the MICE algorithm.
\begin{appendices}
\section{MICE algorithm}
\label{MICE}
In the MICE algorithm, the continuous design space is discretised into a finite grid $\mathbf{X}_{G} \subseteq \mathcal{D}$ such that $\mathbf{X}_{G} = \mathbf{X}_n \cup \mathbf{X}_{cand}$. The latter is a set of candidate points on which the optimisation of the MI criterion is performed. The elements of $\mathbf{X}_{cand}$ are regenerated at each iteration based on a maximin design scheme. The next sample location is obtained by the following optimisation problem 
 \begin{equation}
 	\bx_{n+1} = \underset{\bx \in \mathbf{X}_{cand}}{\arg\!\max}~ s^2_n(\bx) / s^2_{G \backslash (n \cup \bx)}(\bx; \tau^2) ,
 	\label{mice_criter}
 \end{equation}
wherein $G\backslash (n \cup \bx)$ denotes $\mathbf{X}_{G} \backslash \left(\mathbf{X}_n \cup \bx\right)$ and $\tau^2$ is the nugget effect added to the correlation matrix of the GP fitted to $\mathbf{X}_{G} \backslash \left(\mathbf{X}_n \cup \bx\right)$. The inclusion of nugget prevents the denominator of Equation (\ref{mice_criter}) approaching zero. The recommended value for the nugget parameter is one, although it can take any positive value in theory.
\section{Test function expressions}
\label{toy_tests}
The analytic expressions of four test functions used in our experiments are given below.  
\begin{enumerate}
\item $f_1(\bx) = 0.75\exp \left(-\frac{(9x_1 - 2)^2}{4} -\frac{(9x_2 - 2)^2}{4} \right) + 0.75\exp \left(-\frac{(9x_1 + 1)^2}{49}  -\frac{9x_2 + 1)}{10} \right)  \\ ~~~~~~~~~~~
 +  0.5\exp \left(-\frac{(9x_1 - 7)^2}{4} -\frac{(9x_2 - 3)^2}{4} \right) - 0.2 \exp \left( -(9x_1 - 4)^2 - (9x_2 - 7)^2 \right)$.
 \item $f_2(\bx) = - \sum_{i=1}^{4} \boldsymbol{\alpha}_i \exp\left( \sum_{j= 1}^{3} \mathbf{A}_{ij} \left( x_j - \mathbf{P}_{ij} \right)^2 \right)$ where $\boldsymbol{\alpha} = (1, 1.2, 3, 3.2)^\top$, \\
 \[ \mathbf{A} = 
 \begin{bmatrix}
 3    & 10 & 30 \\
 0.1 & 10 & 35 \\
 3    & 10 & 30 \\
 0.1 & 10 & 35
 \end{bmatrix}
 , \quad \mathbf{P} = 10^{-4}
 \begin{bmatrix}
 3689 & 1170  & 2673\\
 4699 & 4387 & 7470 \\
 1091  & 8732 & 5547  \\
 381   &  5743 & 8828
 \end{bmatrix}
 .\]
 \item $f_3(\bx) = 10\sin(\pi x_1x_2) + 20(x_3 - 0.5)^2 +10x_4 + 5x_5$.
 \item $f_4(\bx) = \exp \left( \sin \left( [0.9 (x_1 + 0.48))]^{10} \right)  \right) + x_2x_3 + x_4$.
\end{enumerate}
\section{Real-world function expressions}
\label{real_world_tests}
\paragraph*{OTL circuit function}
The function $f_{OTL}$ is defined as 
\begin{equation*}
f_{OTL}(\bx) = \frac{(V_{b1} + 0.74)\beta(R_{c2} + 9)}{\beta(R_{c2} + 9) + R_f} +  \frac{11.35R_f}{\beta(R_{c2} + 9) + R_f} + \frac{0.74 R_f\beta(R_{c2} + 9)}{(\beta(R_{c2} + 9) + R_f)R_{c1}} ,
\end{equation*}
where $V_{b1} = \frac{12R_{b2}}{R_{b1} + R_{b2}}$. The input variables of $f_{OTL}$ are:
\begin{itemize}
	\item $R_{b1} \in \left[50, 150\right]$ is the resistance $b_1$ (K-Ohms)
	\item $R_{b2} \in \left[25, 70\right]$ is the resistance $b_2$ (K-Ohms)
	\item $R_{f} \in \left[0.5, 3\right]$ is the resistance $f$ (K-Ohms)
	\item $R_{c1} \in \left[1.2, 2.5\right]$ is the resistance $c_1$ (K-Ohms)
	\item $R_{c2} \in \left[0.25, 1.2\right]$ is the resistance $c_2$ (K-Ohms)
	\item $\beta \in \left[50, 300\right]$ is the current gain $c_1$ (Amperes).
\end{itemize}

\paragraph*{Piston simulation function}
The function $f_{piston}$ is defined as 
\begin{align*}
f_{piston}(\bx) = 2\pi \sqrt{\frac{M}{k + S^2 \frac{P_0 V_0 T_a}{T_0 V^2}}} , ~~ \text{where} ~ &V = \frac{S}{2k} \left( \sqrt{A^2 + 4k\frac{P_0V_0}{T_0}T_a} - A\right), \\
& A = P_0S + 19.62M - \frac{kV_0}{S} . 
\end{align*}
The input variables of $f_{piston}$ are:
\begin{itemize}
	\item $M \in \left[30, 60\right]$ is the piston weight (kg)
	\item $S \in \left[0.005, 0.020\right]$ is the piston surface area ($m^2$)
	\item $V_0 \in \left[0.002, 0.010\right]$ is the initial gas volume ($m^3$)
	\item $k \in \left[1000, 5000 \right]$ is the spring coefficient (N/m)
	\item $P_0 \in \left[90000, 110000 \right]$ is the atmospheric pressure (N/$m^2$)
	\item $T_a \in \left[290, 296\right]$ is the ambient temperature (K)
	\item $T_0 \in \left[340, 360 \right]$ is the  filling gas temperature (K).
\end{itemize}
\end{appendices}
\bibliography{biblio}

\begin{thebibliography}{10}

\bibitem{alort2010}
Sylvain Arlot and Alain Celisse.
\newblock {A survey of cross-validation procedures for model selection}.
\newblock {\em Statistics Surveys}, 4:40--79, 2010.

\bibitem{aute2013}
V.~Aute, K.~Saleh, O.~Abdelaziz, S.~Azarm, and R.~Radermacher.
\newblock Cross-validation based single response adaptive design of experiments
  for kriging metamodeling of deterministic computer simulations.
\newblock {\em Structural and Multidisciplinary Optimization}, 48(3):581--605,
  2013.

\bibitem{bachoc2013}
Fran{\c c}ois Bachoc.
\newblock {Cross Validation and maximum likelihood estimation of
  hyper-parameters of Gaussian processes with model misspecification}.
\newblock {\em Computational Statistics and Data Analysis}, 66:55--69, 2013.

\bibitem{banyay2019}
Gregory~A. Banyay, Michael~D. Shields, and John~C. Brigham.
\newblock Efficient global sensitivity analysis for flow-induced vibration of a
  nuclear reactor assembly using kriging surrogates.
\newblock {\em Nuclear Engineering and Design}, 341:1 -- 15, 2019.

\bibitem{beck2016}
J.~Beck and S.~Guillas.
\newblock Sequential design with mutual information for computer experiments
  {(MICE)}: Emulation of a tsunami model.
\newblock {\em SIAM/ASA Journal on Uncertainty Quantification}, 4(1):739--766,
  2016.

\bibitem{benamri2007}
Einat~Neumann Ben-Ari and David~M. Steinberg.
\newblock Modeling data from computer experiments: An empirical comparison of
  kriging with mars and projection pursuit regression.
\newblock {\em Quality Engineering}, 19(4):327--338, 2007.

\bibitem{bensalem2017}
M.~Ben~Salem, O.~Roustant, F.~Gamboa, and L.~Tomaso.
\newblock Universal prediction distribution for surrogate models.
\newblock {\em SIAM/ASA Journal on Uncertainty Quantification},
  5(1):1086--1109, 2017.

\bibitem{box1961}
G.~E.~P. Box and J.~S. Hunter.
\newblock The {2$^{k-p}$} fractional factorial designs part ii.
\newblock {\em Technometrics}, 3(4):449--458, 1961.

\bibitem{brochu2010}
Eric Brochu, Vlad~M. Cora, and Nando de~Freitas.
\newblock {A tutorial on Bayesian optimization of expensive cost functions,
  with application to active user modeling and hierarchical reinforcement
  learning}.
\newblock {\em CoRR}, abs/1012.2599, 2010.

\bibitem{bull2011}
Adam~D. Bull.
\newblock Convergence rates of efficient global optimization algorithms.
\newblock {\em Journal of Machine Learning Research}, 12:2879–2904, 2011.

\bibitem{bukner2020}
Paul-Christian Bürkner, Jonah Gabry, and Aki Vehtari.
\newblock {Approximate leave-future-out cross-validation for Bayesian time
  series models}.
\newblock {\em Journal of Statistical Computation and Simulation},
  90(14):2499--2523, 2020.

\bibitem{dubrule1983}
Olivier Dubrule.
\newblock Cross validation of kriging in a unique neighborhood.
\newblock {\em Journal of the International Association for Mathematical
  Geology}, 15(6):687--699, 1983.

\bibitem{DiceDesign}
Delphine Dupuy, C\'eline Helbert, and Jessica Franco.
\newblock {DiceDesign and DiceEval: two R packages for design and analysis of
  computer experiments}.
\newblock {\em Journal of Statistical Software}, 65(11):1--38, 2015.

\bibitem{friedman1991}
Jerome~H. Friedman.
\newblock Multivariate adaptive regression splines.
\newblock {\em The Annals of Statistics}, 19(1):1--67, 1991.

\bibitem{garud2017}
Sushant~S. Garud, Iftekhar~A. Karimi, and Markus Kraft.
\newblock Design of computer experiments: A review.
\newblock {\em Computers \& Chemical Engineering}, 106:71 -- 95, 2017.
\newblock ESCAPE-26.

\bibitem{BayesianDataAnalysis}
Andrew Gelman, John Carlin, Hal Stern, David Dunson, Aki Vehtari, and Donald
  Rubin.
\newblock {Bayesian Data Analysis, Third Edition (Chapman \& {Hall/CRC} Texts
  in Statistical Science)}, 2013.

\bibitem{ginsbourger2021}
David Ginsbourger and Cedric Schärer.
\newblock {Fast calculation of Gaussian process multiple-fold cross-validation
  residuals and their covariances}, 2021.

\bibitem{gramacy2009}
Robert~B. Gramacy and Herbert K.~H. Lee.
\newblock Adaptive design and analysis of supercomputer experiments.
\newblock {\em Technometrics}, 51(2):130--145, 2009.

\bibitem{legratiet2015}
Loic~Le Gratiet and Claire Cannamela.
\newblock Cokriging-based sequential design strategies using fast
  cross-validation techniques for multi-fidelity computer codes.
\newblock {\em Technometrics}, 57(3):418--427, 2015.

\bibitem{haaland2011}
Ben Haaland and Peter Z.~G. Qian.
\newblock Accurate emulators for large-scale computer experiments.
\newblock {\em The Annals of Statistics}, 39(6):2974--3002, 2011.

\bibitem{jamil2013}
Momin Jamil and Xin-She Yang.
\newblock A literature survey of benchmark functions for global optimization
  problems.
\newblock {\em International Journal of Mathematical Modelling and Numerical
  Optimisation}, 4(2):150--194, 2013.

\bibitem{jin2002}
Ruichen Jin, Wei Chen, and Agus Sudjianto.
\newblock On sequential sampling for global metamodeling in engineering design.
\newblock In {\em Design Engineering Technical Conferences And Computers and
  Information in Engineering}, volume~2, pages 539--548, 2002.

\bibitem{johnson1990}
M.E. Johnson, L.M. Moore, and D.~Ylvisaker.
\newblock Minimax and maximin distance designs.
\newblock {\em Journal of Statistical Planning and Inference}, 26(2):131 --
  148, 1990.

\bibitem{jones2001}
Donald~R. Jones.
\newblock A taxonomy of global optimization methods based on response surfaces.
\newblock {\em Journal of Global Optimization}, 21(4):345--383, Dec 2001.

\bibitem{jones1998}
Donald~R. Jones, Matthias Schonlau, and William~J. Welch.
\newblock Efficient global optimization of expensive black-box functions.
\newblock {\em Journal of Global Optimization}, 13(4):455--492, 1998.

\bibitem{joseph2016}
V.~Roshan Joseph.
\newblock Space-filling designs for computer experiments: A review.
\newblock {\em Quality Engineering}, 28(1):28--35, 2016.

\bibitem{crombecq2011}
Crombecq Karel, Dirk Gorissen, Dirk Deschrijver, and Tom Dhaene.
\newblock A novel hybrid sequential design strategy for global surrogate
  modeling of computer experiments.
\newblock {\em SIAM Journal on Scientific Computing}, 33(4):1948--1974, 2011.

\bibitem{koehler1996}
J.R. Koehler and A.B. Owen.
\newblock Computer experiments.
\newblock In {\em Design and Analysis of Experiments}, volume~13 of {\em
  Handbook of Statistics}, pages 261 -- 308. Elsevier, 1996.

\bibitem{krause2008}
Andreas Krause, Ajit Singh, and Carlos Guestrin.
\newblock Near-optimal sensor placements in {G}aussian processes: Theory,
  efficient algorithms and empirical studies.
\newblock {\em Journal of Machine Learning Research}, 9:235--284, February
  2008.

\bibitem{lam2008}
Chen~Quin Lam.
\newblock {\em Sequential adaptive designs in computer experiments for response
  surface model fit}.
\newblock PhD thesis, Columbus, OH, USA, 2008.
\newblock AAI3321369.

\bibitem{legratiet2012}
Loic {Le Gratiet} and Claire {Cannamela}.
\newblock {Kriging-based sequential design strategies using fast
  cross-validation techniques with extensions to multi-fidelity computer
  codes}.
\newblock {\em arXiv e-prints}, page arXiv:1210.6187, 2012.

\bibitem{li2009}
Genzi Li, Vikrant Aute, and Shapour Azarm.
\newblock An accumulative error based adaptive design of experiments for
  offline metamodeling.
\newblock {\em Structural and Multidisciplinary Optimization}, 40(1):137--157,
  2009.

\bibitem{liang2014}
Haoquan Liang, Ming Zhu, and Zhe Wu.
\newblock Using cross-validation to design trend function in kriging surrogate
  modeling.
\newblock {\em AIAA Journal}, 52(10):2313--2327, 2014.

\bibitem{liu2017}
D.~Liu, A.~Litvinenko, C.~Schillings, and V.~Schulz.
\newblock Quantification of airfoil geometry-induced aerodynamic
  uncertainties--comparison of approaches.
\newblock {\em SIAM/ASA Journal on Uncertainty Quantification}, 5(1):334--352,
  2017.

\bibitem{liu2018}
Haitao Liu, Yew-Soon Ong, and Jianfei Cai.
\newblock A survey of adaptive sampling for global metamodeling in support of
  simulation-based complex engineering design.
\newblock {\em Structural and Multidisciplinary Optimization}, 57(1):393--416,
  2018.

\bibitem{liu2015}
Haitao Liu, Shengli Xu, Ying Ma, Xudong Chen, and Xiaofang Wang.
\newblock An adaptive {B}ayesian sequential sampling approach for global
  metamodeling.
\newblock {\em Journal of Mechanical Design}, 138(1), 2015.

\bibitem{maatouk2015}
Hassan Maatouk, Olivier Roustant, and Yann Richet.
\newblock Cross-validation estimations of hyper-parameters of {G}aussian
  processes with inequality constraints.
\newblock {\em Procedia Environmental Sciences}, 27:38--44, 2015.
\newblock Spatial Statistics conference 2015.

\bibitem{martin2002}
Jay Martin and Timothy Simpson.
\newblock Use of adaptive metamodeling for design optimization.
\newblock In {\em 9th AIAA/ISSMO Symposium on Multidisciplinary Analysis and
  Optimization}, pages 1--9, 2002.

\bibitem{martino2017}
Luca Martino, Valero Laparra, and Gustau Camps-Valls.
\newblock {Probabilistic cross-validation estimators for Gaussian process
  regression}.
\newblock In {\em 25th European Signal Processing Conference (EUSIPCO)}, pages
  823--827, 2017.

\bibitem{mckay1979}
M.~D. McKay, R.~J. Beckman, and W.~J. Conover.
\newblock A comparison of three methods for selecting values of input variables
  in the analysis of output from a computer code.
\newblock {\em Technometrics}, 21(2):239--245, 1979.

\bibitem{mohammadi2016}
Hossein Mohammadi.
\newblock {\em Kriging-based black-box global optimization: analysis and new
  algorithms}.
\newblock PhD thesis, \'Ecole Nationale Sup\'erieure des Mines de
  Saint-\'Etienne, France, 2016.

\bibitem{mullen2011}
Katharine~M. Mullen, David Ardia, David~L. Gil, Donald Windover, and James
  Cline.
\newblock {DEoptim: An R package for global optimization by differential
  evolution}.
\newblock {\em Journal of Statistical Software, Articles}, 40(6):1--26, 2011.

\bibitem{neal1998}
Radford~M. Neal.
\newblock Regression and classification using {G}aussian process priors.
\newblock pages 475--501. Bayesian {S}tatistics 6, Oxford University Press,
  1998.

\bibitem{owen1992}
Art~B. Owen.
\newblock Orthogonal arrays for computer experiments, integration and
  visualization.
\newblock {\em Statistica Sinica}, 2(2):439--452, 1992.

\bibitem{picheny2010}
V.~Picheny, D.~Ginsbourger, O.~Roustant, R.~T. Haftka, and N.~H. Kim.
\newblock Adaptive designs of experiments for accurate approximation of a
  target region.
\newblock {\em Journal of Mechanical Design}, 132(7):1--9, 2010.

\bibitem{ponweiser2008}
W.~{Ponweiser}, T.~{Wagner}, and M.~{Vincze}.
\newblock Clustered multiple generalized expected improvement: A novel infill
  sampling criterion for surrogate models.
\newblock In {\em 2008 IEEE Congress on Evolutionary Computation (IEEE World
  Congress on Computational Intelligence)}, pages 3515--3522, 2008.

\bibitem{pronzato2012}
Luc Pronzato and Werner~G. M{\"u}ller.
\newblock Design of computer experiments: space filling and beyond.
\newblock {\em Statistics and Computing}, 22(3):681--701, 2012.

\bibitem{GPML}
Carl~Edward Rasmussen and Christopher K.~I. Williams.
\newblock {\em Gaussian processes for machine learning (adaptive computation
  and machine learning)}.
\newblock The MIT Press, 2005.

\bibitem{roustant2012}
Olivier Roustant, David Ginsbourger, and Yves Deville.
\newblock {DiceKriging}, {DiceOptim}: Two {R} packages for the analysis of
  computer experiments by kriging-based metamodeling and optimization.
\newblock {\em Journal of Statistical Software}, 51(1):1--55, 2012.

\bibitem{ryzhov2016}
Ilya Ryzhov.
\newblock On the convergence rates of expected improvement methods.
\newblock {\em Operations Research}, 64(6), 2016.

\bibitem{sacks1989}
Jerome Sacks, William~J. Welch, Toby~J. Mitchell, and Henry~P. Wynn.
\newblock Design and analysis of computer experiments.
\newblock {\em Statistical Science}, 4(4):409--423, 1989.

\bibitem{santner2003}
T.~J. Santner, Williams B., and Notz W.
\newblock {\em The design and analysis of computer experiments}.
\newblock Springer-Verlag, 2003.

\bibitem{schonlau1997}
Matthias Schonlau.
\newblock {\em Computer experiments and global optimization}.
\newblock PhD thesis, University of Waterloo, 1997.

\bibitem{sheikholeslami2017}
Razi Sheikholeslami and Saman Razavi.
\newblock Progressive latin hypercube sampling: an efficient approach for
  robust sampling-based analysis of environmental models.
\newblock {\em Environmental Modelling \& Software}, 93:109 -- 126, 2017.

\bibitem{shewry1987}
M.~C. Shewry and H.~P. Wynn.
\newblock Maximum entropy sampling.
\newblock {\em Journal of Applied Statistics}, 14(2):165--170, 1987.

\bibitem{simpson2001}
T.W. Simpson, J.D. Poplinski, P.~N. Koch, and J.K. Allen.
\newblock Metamodels for computer-based engineering design: Survey and
  recommendations.
\newblock {\em Engineering with Computers}, 17(2):129--150, 2001.

\bibitem{storn1997}
Rainer Storn and Kenneth Price.
\newblock Differential evolution -- a simple and efficient heuristic for global
  optimization over continuous spaces.
\newblock {\em Journal of Global Optimization}, 11(4):341--359, 1997.

\bibitem{vazquez2010}
Emmanuel Vazquez and Julien Bect.
\newblock Convergence properties of the expected improvement algorithm with
  fixed mean and covariance functions.
\newblock {\em Journal of Statistical Planning and Inference},
  140(11):3088--3095, 2010.

\bibitem{vernon2018}
Ian Vernon, Junli Liu, Michael Goldstein, James Rowe, Jen Topping, and Keith
  Lindsey.
\newblock Bayesian uncertainty analysis for complex systems biology models:
  emulation, global parameter searches and evaluation of gene functions.
\newblock {\em BMC Systems Biology}, 12(1):1, 2018.

\bibitem{viana2009}
Felipe A.~C. Viana, Raphael~T. Haftka, and Valder Steffen.
\newblock Multiple surrogates: how cross-validation errors can help us to
  obtain the best predictor.
\newblock {\em Structural and Multidisciplinary Optimization}, 39(4):439--457,
  2009.

\bibitem{volodina2020}
Victoria Volodina and Daniel Williamson.
\newblock Diagnostics-driven nonstationary emulators using kernel mixtures.
\newblock {\em SIAM/ASA Journal on Uncertainty Quantification}, 8(1):1--26,
  2020.

\bibitem{williamson2015}
Daniel Williamson.
\newblock {Exploratory ensemble designs for environmental models using
  k-extended Latin Hypercubes}.
\newblock {\em Environmetrics}, 26(4):268--283, 2015.

\bibitem{youngmok2014}
Youngmok Yun, Hyun-Chul Kim, Sung~Yul Shin, Junwon Lee, Ashish~D. Deshpande,
  and Changhwan Kim.
\newblock {Statistical method for prediction of gait kinematics with Gaussian
  process regression}.
\newblock {\em Journal of Biomechanics}, 47(1):186--192, 2014.

\bibitem{zhan2017}
Dawei Zhan, Jiachang Qian, and Yuansheng Cheng.
\newblock {Pseudo expected improvement criterion for parallel EGO algorithm}.
\newblock {\em Journal of Global Optimization}, 68(3):641--662, 2017.

\bibitem{zhang2015}
Yongli Zhang and Yuhong Yang.
\newblock Cross-validation for selecting a model selection procedure.
\newblock {\em Journal of Econometrics}, 187(1):95--112, 2015.

\end{thebibliography}
\bibliographystyle{plain}
\end{document}